\documentclass[aps,twocolumn,prc,superscriptaddress,preprintnumbers,nofootinbib,floatfix,amssymb,amsfonts,amsmath]{revtex4-2}
\usepackage{graphicx}
\usepackage{epsf}
\usepackage{amsmath,amssymb}
\usepackage{bm}
\usepackage{color}
\usepackage{ulem}

\newcommand{\nuc}[2]{{}^{#2} \mathrm{#1}}
\begin{document}
\title{
  Persistence of cluster structure in the ground state of $\nuc{B}{11}$
}%
\author{Naoyuki Itagaki}
\affiliation{
  Yukawa Institute for Theoretical Physics, Kyoto University,
  Kitashirakawa Oiwake-Cho, Kyoto 606-8502, Japan
}
\author{Tomoya Naito}
\affiliation{
  Department of Physics, Graduate School of Science, The University of Tokyo, Tokyo 113-0033, Japan}
\affiliation{
  RIKEN Nishina Center, Wako 351-0198, Japan
}
\author{Yuichi Hirata}
\affiliation{
  Ninomiya, Tsukuba, Ibaraki 305-0051, Japan
}
\preprint{RIKEN-QHP-503}
\preprint{RIKEN-iTHEMS-Report-21}
\date{\today}
\begin{abstract}
  The persistence of $\alpha+\alpha+t$ cluster configuration in the ground state 
  of $\nuc{B}{11}$ is discussed.
  The $\alpha$ particles 
  emitted from excited states of $\nuc{C}{12}$
  could be utilized for cancer treatment,
  where the cluster states
  are created by a clinical proton beam on the $\nuc{B}{11}$ target.
  Although the cross section has been already available,
  for the nuclear structure side,
  whether the ground state of $\nuc{B}{11}$ contains the seeds of the cluster states 
  is a crucial question,
  since the cluster structure may be washed out by
  the spin-orbit interaction, which has been
  known as a driving force to break the
  $\alpha$ clusters.
  For this purpose, 
  in addition to the basis states with cluster configurations, we include 
  their breaking effects by employing
  the antisymmetrized quasi cluster model (AQCM).
  The inclusion of the breaking effect of 
  $\alpha+\alpha+t$ cluster structure is found to contribute to the lowering of the 
  ground-state energy by about $ 2 \, \mathrm{MeV} $.
  Assuming the typical three-$\alpha$ cluster state of $\nuc{C}{12}$ 
  as an equilateral triangular configuration with the relative distances of $\approx 4 \, \mathrm{fm}$,
  the ground state of $\nuc{B}{11}$ is found to have a certain squared overlap 
  with such state when a proton approaches.
  The third $3/2^{-}$ state
  has been suggested as a cluster state both theoretically and experimentally,
  and we confirmed the well-developed clustering.
\end{abstract}
\maketitle
%
\section{Introduction}
\par
In describing the structure of
light nuclei, the $\nuc{He}{4}$ nucleus is often treated as 
a subsystem called $\alpha$ cluster due to its large binding energy
in the free space.
The most famous example of the state comprised of $\alpha$ clusters is the second $0^{+}$ state of $\nuc{C}{12}$
called the Hoyle state, which is a three $\alpha$ state with large relative
distances between $\alpha$ clusters~\cite{Hoyle1954,FREER20141}.
The cluster models have been found to be quite effective in describing various properties of such state~\cite{PTPS.68.29,PhysRevLett.87.192501}.
\par
Cluster states of $\nuc{C}{12}$, which are located higher than Hoyle state 
in energy, are recently utilized for cancer treatment. 
A clinical proton beam
on the $\nuc{B}{11}$ target induces excited states of $\nuc{C}{12}$ with three $\alpha $ configurations
above the three-$\alpha$ threshold energy and emitted $\alpha$ particles from these states
as their decay are used for destroying cancer cells~\cite{SciRep8-1141}.
For $\nuc{C}{12}$, the $\nuc{B}{11}+p$ threshold corresponds to the excitation energy of $ 16.0 \, \mathrm{MeV} $,
well above three $\alpha$ threshold energy of $ 7.4 \, \mathrm{MeV} $,
but the neutron threshold ($ 18.7 \, \mathrm{MeV} $) is not opened.
Therefore, the excited states decay by emitting $\alpha$ particles. 
This method is called Proton Boron Capture Therapy (PBCT).
The three-$\alpha$ cluster states of $\nuc{C}{12}$ in this energy region
have been investigated for decades~\cite{PLB696.26.Stave}, and the data of $p$--$\nuc{B}{11}$ scattering
cross section has been accumulated.
\par
For the nuclear structure side, it would be still intriguing to investigate whether $\alpha+\alpha+t$ structure persists 
in the ground state of $\nuc{B}{11}$. 
The fact that this method work means that the ground state of $\nuc{B}{11}$ contains 
enough component of the $\alpha+\alpha+t$ cluster
configurations, which is the seed of the three $\alpha$ states of $\nuc{C}{12}$.
It has been discussed in $\nuc{B}{11}$ that
the third $3/2^{-}$ state is a candidate for the state with a cluster structure~\cite{KAWABATA20076},
but it is considered that in the ground state, the cluster structure is washed out to some extent
as in the case of $\nuc{C}{12}$.
Indeed, the spin-orbit interaction is 
known to work as a driving force to break the
$\alpha$ clusters~\cite{PhysRevC.70.054307}.
\par
In most of the conventional $\alpha$ cluster models,  
the contribution of the non-central interactions (spin-orbit and tensor interactions) vanishes.
If non-central interaction acts attractively by incorporating shell-model states
in the model space,
we must extend the model space and break the cluster structure. 
As well known, the spin-orbit  interaction
is important in the shell model;
the observed magic numbers of $28$, $50$, and $126$
correspond to the subclosure configurations of 
$f_{7/2}$, $g_{9/2}$, and $h_{11/2}$
of the $jj$-coupling shell model~\cite{Mayer}.
\par
This spin-orbit contribution is included
by extending the cluster model;
we have developed the antisymmetrized quasi cluster model
(AQCM)~\cite{PhysRevC.71.064307,PhysRevC.75.054309,PhysRevC.79.034308,PhysRevC.83.014302,PhysRevC.87.054334,ptep093D01,ptep063D01,ptepptx161,PhysRevC.94.064324,PhysRevC.97.014307,PhysRevC.98.044306,PhysRevC.101.034304,PhysRevC.102.024332,PhysRevC.103.044303}.
This method allows us to smoothly transform $\alpha$-cluster model wave functions to
$jj$-coupling shell model ones, and
we call the clusters that feel the effect of the spin-orbit interaction owing to this model quasi clusters.
We have previously introduced AQCM to $\nuc{C}{12}$ and discussed the
competition between the cluster states and $jj$-coupling shell model state~\cite{PhysRevC.94.064324}.
The consistent description of $\nuc{C}{12}$ and $\nuc{O}{16}$, which has been a long-standing problem
of microscopic cluster models, has been achieved.
Also, not only the competition between the
cluster states and the lowest shell-model configuration,
the effect of single-particle excitation was further included 
for the description of the ground state~\cite{PhysRevC.103.044303}.
\par
For $\nuc{B}{11}$,
until now, various cluster models~\cite{Nishioka-PTP.62.424,DESCOUVEMONT1995532,PhysRevC.82.064315,PhysRevC.98.054323}
and antisymmetrized molecular dynamics (AMD)~\cite{PhysRevC.75.024302,PhysRevC.85.054320} have been applied,
where the main focus was the clustering in the excited states.
In this paper, however, we couple cluster model space and shell-model one and
investigate the persistence of $\alpha+\alpha+t$ cluster configuration 
in the ground state  of $\nuc{B}{11}$.
In addition to the basis states with cluster configurations, cluster breaking ones are prepared with AQCM,
where the contribution of the spin-orbit interaction plays an important role.
Furthermore, the basis states with the $\alpha+\alpha+3N$ configurations are generated.
All of these basis states are superposed based on
the framework of the generator coordinate method (GCM),
and cluster-shell competition is microscopically investigated.
The interactions used are the same as those in our previous analysis on $\nuc{C}{12}$~\cite{PhysRevC.103.044303}.
For the central part, the Tohsaki interaction~\cite{PhysRevC.49.1814},
which has finite range three-body terms, is adopted.
There is no free parameter to be adjusted.
For the spin-orbit part, we use the spin-orbit term of the G3RS interaction~\cite{PTP.39.91},
whose strength is set to give
consistent description of $\nuc{C}{12}$ and $\nuc{O}{16}$~\cite{PhysRevC.94.064324}.
\par
This paper is organized as follows. 
The framework is described in  Sec.~\ref{Frame}.
The results are shown in Sec.~\ref{Results}.
The conclusions are presented in Sec.~\ref{Concl}.
%
\section{framework}
\label{Frame}
%
\par
\begin{figure}[b]
  \centering
  \includegraphics[width=1.0\linewidth]{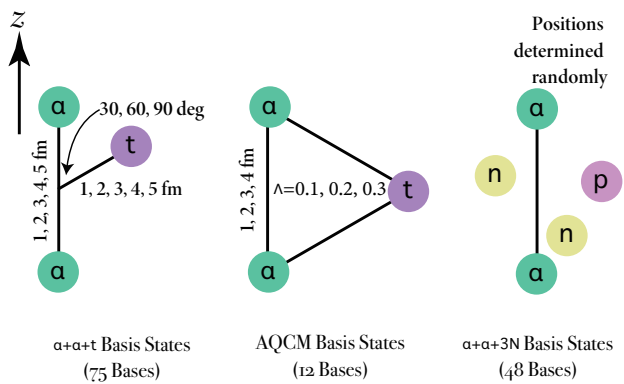} 
  \caption{
    Schematic figure of basis sets used in this work.
    See the text for more detail.}
  \label{bases}
\end{figure}

The wave function is fully antisymmetrized, and
we superpose three kinds of the basis states:
$\alpha+\alpha+t$ basis states (75 bases),
AQCM basis states (12 bases),
and $\alpha+\alpha+3N$ basis states (48 bases),
whose schematic figure is shown in Fig.~\ref{bases}.
All these states are superposed based on
the GCM
after the
angular momentum projection and amplitude for each basis state is determined by diagonalizing the norm and Hamiltonian matrices.  
%
%
\subsection{Single-particle wave function}
\par
In our framework, each single particle is described by a Gaussian form
as in many other cluster models including the Brink model~\cite{Brink},
\begin{equation}	
  \phi^{\tau, \sigma} \left( \bm{r} \right)
  =
  \left(  \frac{2\nu}{\pi} \right)^{\frac{3}{4}} 
  \exp \left[- \nu \left(\bm{r} - \bm{\zeta} \right)^{2} \right] \chi^{\tau,\sigma}, 
  \label{spwf} 
\end{equation}
where the Gaussian center parameter $\bm{\zeta}$
is related to the expectation 
value of the position of the nucleon,
and $\chi^{\tau,\sigma}$ is the spin-isospin part of the wave function.
For the size parameter $\nu$, 
here we use $\nu = 0.23 \, \mathrm{fm}^{-2}$, which gives the optimal $0^{+}$ energy
of $\nuc{C}{12}$ within a single AQCM basis state.
The Slater determinant is constructed from 
these single-particle wave functions by antisymmetrizing them.
\subsection{Basis states of $\alpha+\alpha+t$ cluster configurations}
\par
When four single-particle 
wave functions with different spin and isospin
share a common 
$\bm{\zeta}$ value,  an $\alpha$ cluster is formed.
Similarly, when two neutrons with opposite spin orientations and a proton share the same
$\bm{\zeta}$, a triton cluster is formed. We prepare 75 basis states with
different $\alpha+\alpha+t$ configurations.
The distance between two $\alpha$ clusters is changed from $ 1 \, \mathrm{fm} $ to $ 5 \, \mathrm{fm} $ in the step of $ 1 \, \mathrm{fm} $.
The triton is set at the places with the distance of $ 1 $--$5\,\mathrm{fm}$ 
from the center of two $\alpha$ clusters, which is changed in the step of $1\,\mathrm{fm}$,
and the  
axis measured from the $\alpha$--$\alpha$ axis is set to $30^\circ$, $60^\circ$, and $90^\circ$.
In all these states,
the $\bm{\zeta}$ values in Eq.~\eqref{spwf} are real numbers.
\subsection{Basis states of AQCM}
\par
This cluster wave function is transformed into
$jj$-coupling shell model based on the AQCM,
by which the contribution of the spin-orbit interaction
due to the breaking of $\alpha$ clusters is included.
Here
the $\bm{\zeta}$ values in Eq.~\eqref{spwf} are changed to complex numbers.
When the original value of the Gaussian center parameter $\bm{\zeta}$
is $\bm{R}$,
which is 
real and
related to the spatial position of this nucleon, 
it is transformed 
by adding the imaginary part as
\begin{equation}
  \bm{\zeta} = \bm{R} + i \Lambda \bm{e}^{\text{spin}} \times \bm{R}, 
  \label{AQCM}
\end{equation}
where $\bm{e}^{\text{spin}}$ is a unit vector for the intrinsic-spin orientation of this
nucleon. 
The control parameter $\Lambda$ is associated with the breaking of the cluster,
and two nucleons with opposite spin orientation have $\bm{\zeta}$ values
that are complex conjugate to each other.
This situation corresponds to the time-reversal motion of two nucleons.
After this transformation, the $\alpha$ clusters are called quasi clusters.
\par
In our previous analysis on $\nuc{C}{12}$, we have prepared three quasi clusters
with an equilateral triangular shape.
We introduced two parameters, $R$ representing the $\alpha$--$\alpha$ distances
and $\Lambda$ for the breaking of the $\alpha$ clusters.
The
subclosure configuration of 
$ \left( s_{1/2} \right)^2 \, \left( p_{3/2} \right)^4$
of the $jj$-coupling shell model can be obtained at the limit of
$R \to 0$ and $\Lambda = 1$.
\par
For $\nuc{B}{11}$, we remove one spin-down proton from a quasi cluster.
The $R$ and $\Lambda$ values are taken as $R = 1$, $2$, $3$, $4 \, \mathrm{fm} $ and $\Lambda = 0.1 $, $0.2$, $0.3$
($\Lambda = 0$ states are not necessary because they have large overlaps with the $\alpha+\alpha+t$ basis states).
\par
\begin{figure}[tb]
  \centering
  \includegraphics[width=5.5cm]{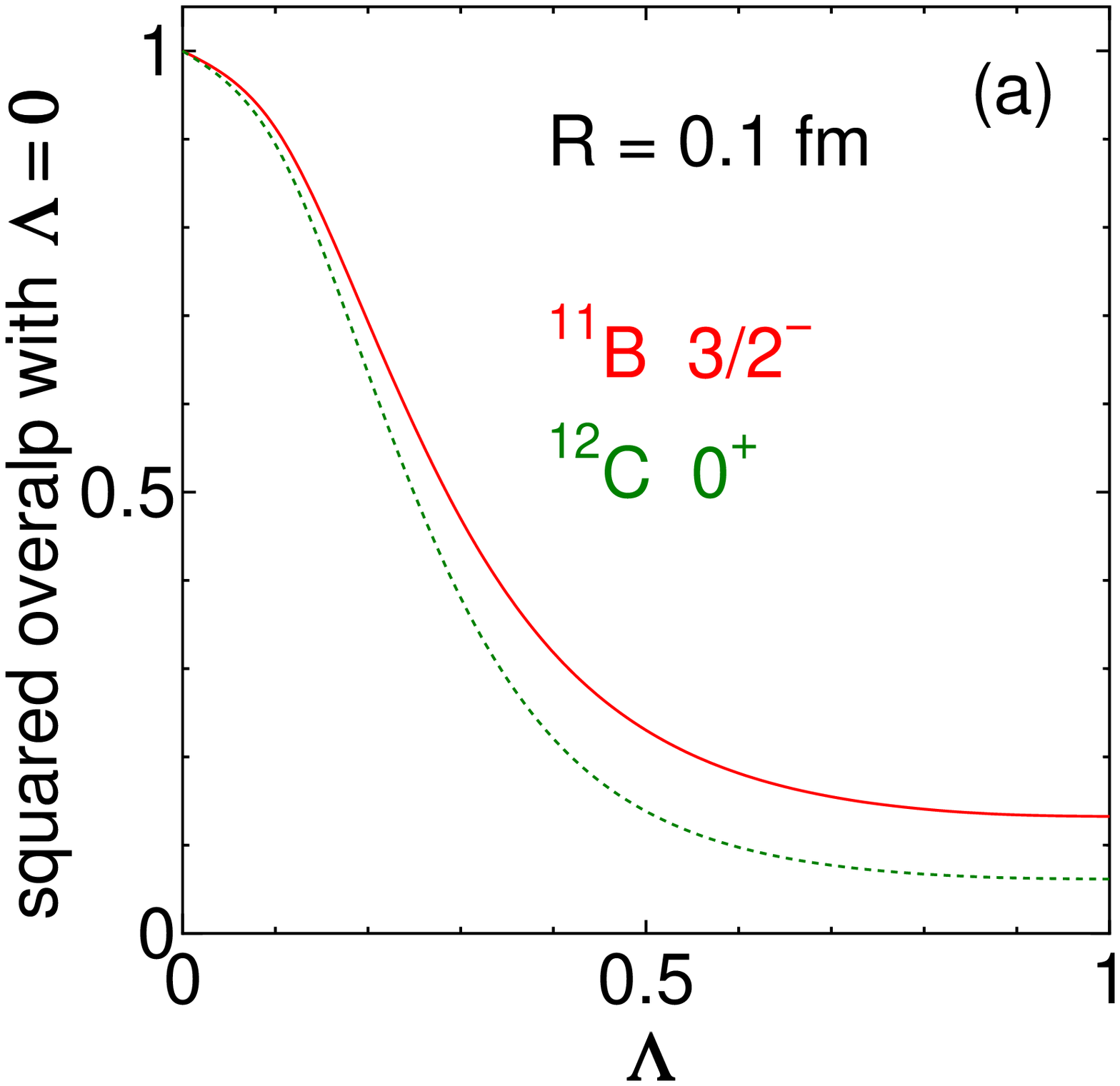} 
  \includegraphics[width=5.5cm]{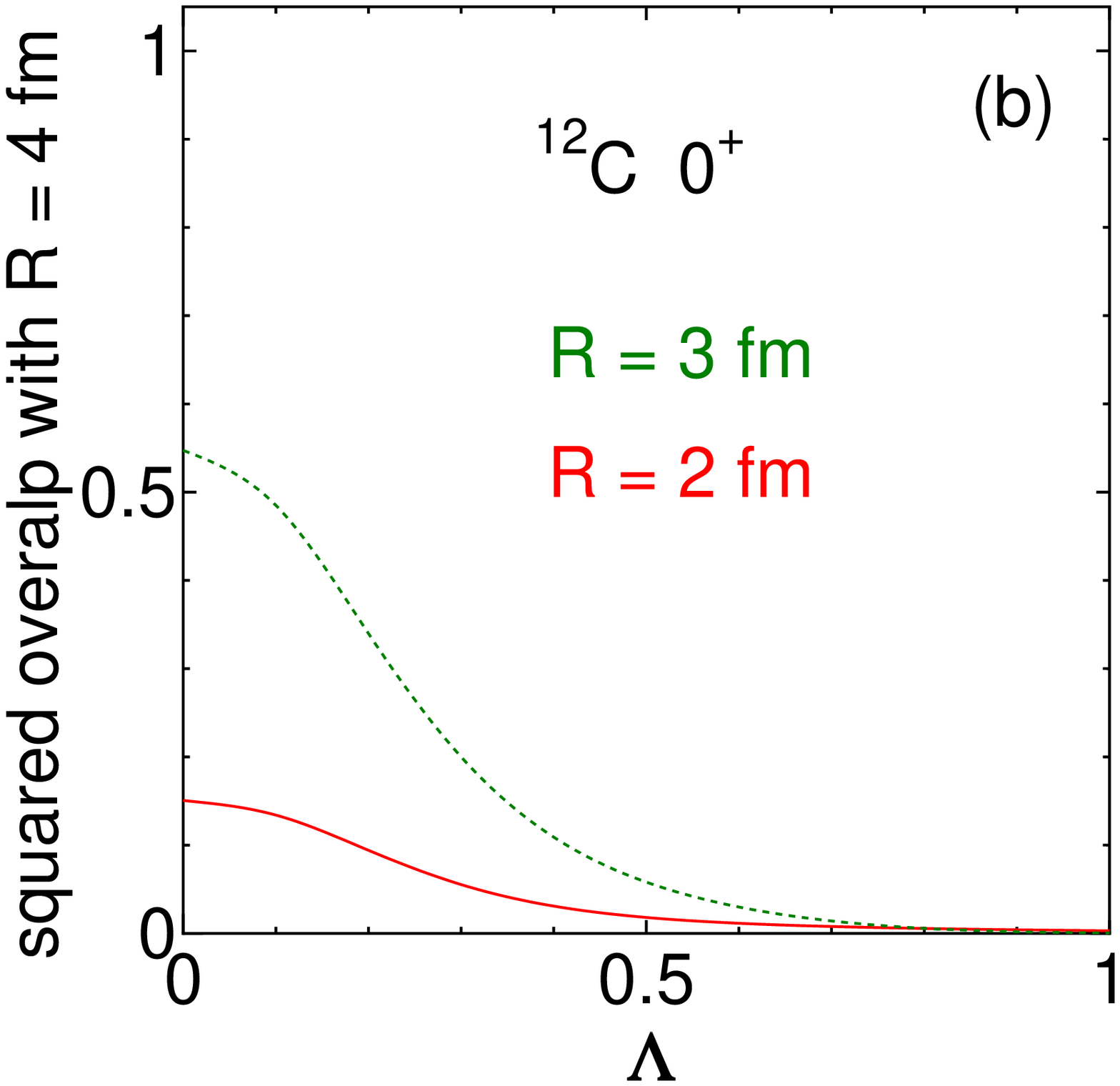} 
  \caption{
    (a) Squared overlap between the three-dimensional harmonic oscillator (shell-model) and AQCM
    as functions of the $\Lambda$ value.
    The dotted line is for the $0^{+}$ state of $\nuc{C}{12}$ between equilateral triangular configuration and very small
    distances between three quasi clusters
    and finite $\Lambda$ states.
    The solid line is for the $3/2^{-}$ of $\nuc{B}{11}$.
    (b). Squared overlaps between a typical three-$\alpha$ cluster state 
    ($R = 4 \, \mathrm{fm} $)
    and AQCM basis states.
    The solid and dotted lines are for the squared overlap with this typical cluster state
    and the AQCM basis stats with $R = 2 \, \mathrm{fm} $ and $R = 3 \, \mathrm{fm} $ as functions of $\Lambda$, respectively.}
  \label{c12-sqovlp}
\end{figure}

It has been discussed that the cluster model space covers the model space of the shell model,
if the wave function is antisymmetrized and we take the small-distance limit between clusters.
This statement is correct 
when the term of the ``shell model'' is used
in the sense of the three-dimensional harmonic oscillator.
The cluster-model wave function becomes the three-dimensional harmonic oscillator
one at the limit of the small relative distances.
However, 
the three-dimensional harmonic oscillator wave function
(corresponding to $ \Lambda = 0 $ in AQCM) 
is quite different from the $jj$-coupling shell-model one
(corresponding to $ \Lambda = 1 $ in AQCM),
as we can easily assess using our AQCM.
Figure~\ref{c12-sqovlp}(a) shows the squared overlap between the three-dimensional harmonic oscillator (cluster model with small distances) and AQCM
as functions of the $\Lambda$ value.
The dotted line is for $\nuc{C}{12}$ with equilateral triangular configuration and very small
distances between three quasi clusters. The vertical axis shows the squared overlap
between $\Lambda=0$ and finite $\Lambda$ states. The angular momentum and parity are projected to $0^{+}$.
The dotted line rapidly drops and 
the state with $\Lambda = 0$
(identical to the three-dimensional harmonic oscillator one)
has very small squared overlap 
with the subclosure configuration of the $jj$-coupling shell model
($\left(s_{1/2}\right)^4\,\left(p_{3/2}\right)^8$)
of about $ 5 \, \% $
at $\Lambda = 1$.
The solid line is for the $3/2^{-}$ of $\nuc{B}{11}$.
Since one proton is missing compared with $\nuc{C}{12}$,
the squared overlap between the three-dimensional harmonic oscillator and $jj$-coupling shell model
increases, but the value is still quite small (slightly above $ 10 \, \% $).
\par
Next, the squared overlaps between a typical three-$\alpha$ cluster state and AQCM basis states are shown in 
Fig.~\ref{c12-sqovlp}(b).
Here, the typical three-$\alpha$ cluster state means an equilateral triangular configuration
with the $\alpha$--$\alpha$ distance ($R$) of $ 4\,\mathrm{fm}$.
The solid and dotted lines are for the squared overlap with this typical cluster state
and the AQCM basis states with $R = 2 \, \mathrm{fm} $ and $R = 3 \, \mathrm{fm} $ as functions of $\Lambda$, respectively.
In our previous work, we have discussed that $R = 2.0 \, \mathrm{fm} $ and $\Lambda = 0.2$ is the optimal AQCM basis state,
which is found have the squared overlap of $10\,\%$ with the typical cluster state.
%
%
\subsection{Basis states of $\alpha+\alpha+3N$}
\par
To describe the single-particle nature of three nucleons outside two $\alpha$ clusters,
we prepare $\alpha+\alpha+3N$ basis states,
where $\alpha$--$\alpha$ distance is determined using the random number
(between $0$--$5\,\mathrm{fm}$ with equal probability) and positions of three nucleons 
(a spin-up proton and two neutrons with opposite spin directions)
are also
determined randomly. 
We generate 48 basis states and include them
in the diagonalization process of the norm and Hamiltonian matrices.
\subsection{$J^\pi$ projection and GCM}
\par
As we have mentioned before, totally we generate 135 intrinsic states. 
The basis states from 1 to 75 are $\alpha+\alpha+t$ states with various configurations,
those from 76 to 87 are AQCM basis states with
$R = 1 $, $2$, $3$, $4 \, \mathrm{fm} $ and $\Lambda = 0.1, 0.2, 0.3$,
and those from 88 to 135 are $\alpha+\alpha+3N$ configurations
with the $\alpha$--$\alpha$ distance of $0$--$5 \, \mathrm{fm}$.
These 135 basis states are numerically projected to eigenstates of angular momentum and parity.
After the angular momentum projection, different $K$ number states are generated from 
the same intrinsic basis state. 
Here, the $z$ direction of the intrinsic frame is parallel to the axis of two $\alpha$ (quasi) clusters.
These different $K$ states are treated independently 
when we determine the coefficients for the linear combination of the basis states
based on GCM.
Therefore, after the angular momentum projection,
the number of the basis states increases to 540,
the basis states of 1--135, 136--270, 271--405, and 406--540 are
$K=1/2$, $K=3/2$, $K=-1/2$, and $K=-3/2$, respectively.

\subsection{Hamiltonian}
\par
The Hamiltonian consists of the kinetic energy and 
potential energy terms.
For the potential part, the interaction consists of the central, spin-orbit, and Coulomb terms. 
The interactions are the same as those in our previous analysis on $\nuc{C}{12}$~\cite{PhysRevC.103.044303}.
For the central part, the Tohsaki interaction~\cite{PhysRevC.49.1814} is adopted.
This interaction has finite-range three-body terms in addition to two-body terms,
which is designed to reproduce both saturation properties
and scattering phase shifts of two $\alpha$ clusters.
For the spin-orbit part, 
There is no free parameter left for each nucleus for this interaction.
For the spin-orbit part, we use the spin-orbit term of the G3RS interaction~\cite{PTP.39.91},
which is a realistic interaction originally developed to reproduce the nucleon-nucleon scattering phase shifts.
The strength of the spin-orbit interactions is set to $V_{ls}^1=V_{ls}^2=1800 \, \mathrm{MeV}$,
which allows consistent description of $\nuc{C}{12}$ and $\nuc{O}{16}$~\cite{PhysRevC.94.064324}.
\section{Results}
\label{Results}
\subsection{Energy levels of $\nuc{B}{11}$}
\par
\begin{figure}[tb]
  \centering
  \includegraphics[width=5.5cm]{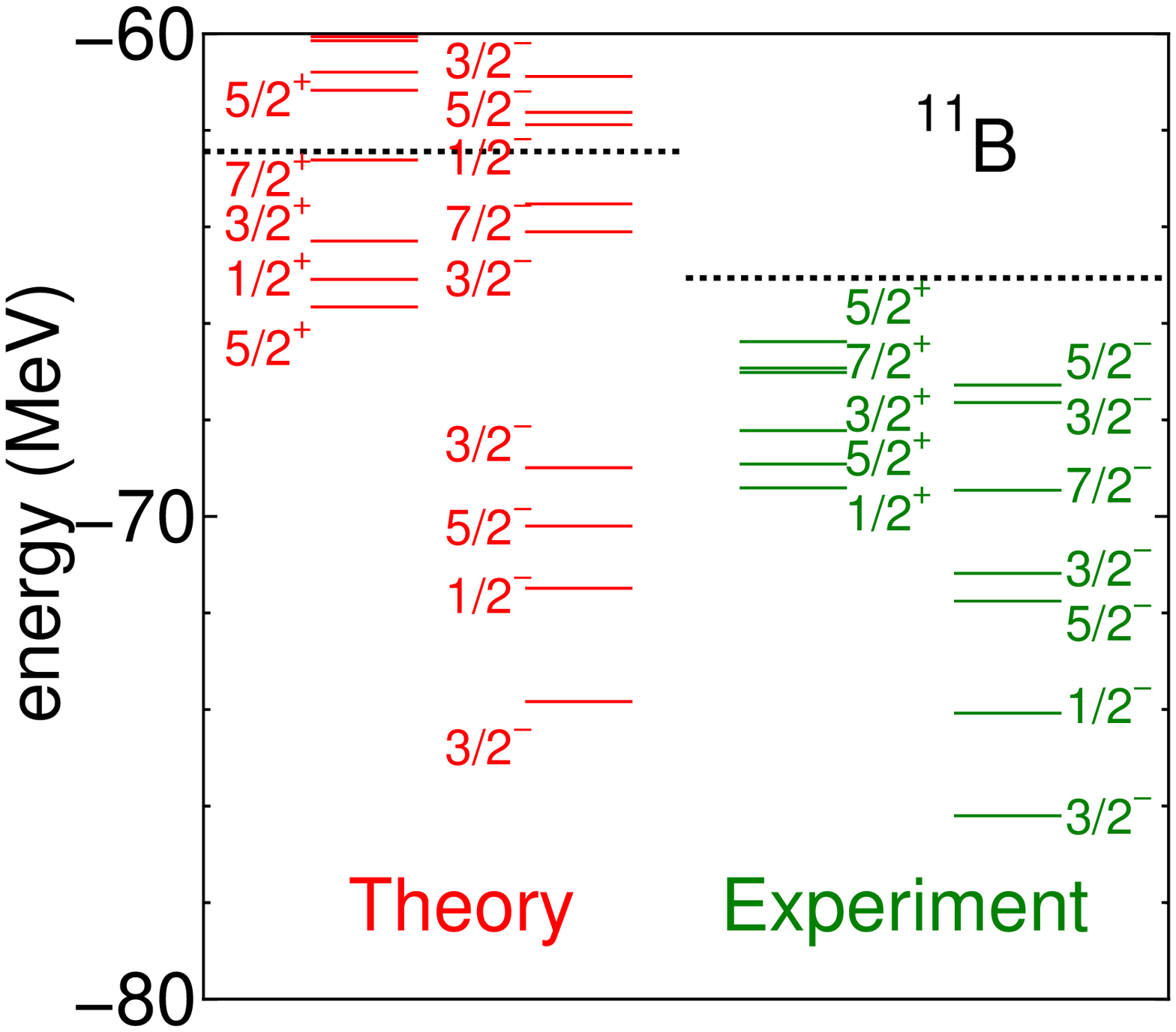} 
  \caption{
    Energy levels of $\nuc{B}{11}$. 
    The 135 basis states are projected to the eigenstates of angular momentum and parity,
    and basis states with $K=1/2$, $3/2$, $-1/2$, and $-3/2$ are independently treated
    when diagonalizing the norm and Hamiltonian matrices based on GCM.
    Dotted lines are threshold energies of  $\alpha+\alpha+t$.}
  \label{b11-level}
\end{figure}

We project 135 basis states to the eigenstates of angular momentum and parity,
and basis states with $K=1/2$, $3/2$, $-1/2$, and $-3/2$ are independently treated.
Thus, based on GCM, we diagonalize the norm and Hamiltonian matrices with the dimension of 540.
The obtained energy levels of $\nuc{B}{11}$ are listed in Fig.~\ref{b11-level},
where dotted lines are threshold energies of  $\alpha+\alpha+t$.
The ground $3/2^{-}$ state is obtained at $-73.84 \, \mathrm{MeV} $,
compared with the experimental value of $-76.20 \, \mathrm{MeV} $.
Although we have no adjustable parameter for the central interaction
and the spin-orbit strength is determined in our previous work for $\nuc{C}{12}$,
the absolute value of the calculated $\nuc{B}{11}$ energy is quite reasonable.
\par
The theoretical binding energy underestimates the experimental one by $2.4\,\mathrm{MeV}$, 
but the agreement between theoretical and experimental binding energies becomes even better
if we measure them from the $\alpha+\alpha+t$ threshold.
This is because the underestimate can be absorbed in the internal energies of the clusters;
the theoretical value of $\alpha$ (triton) energy is $-27.31 \, \mathrm{MeV} $ ($-7.82 \, \mathrm{MeV} $),
compared with the experimental value of $-28.29 \, \mathrm{MeV} $ ($-8.48 \, \mathrm{MeV} $). 
The ground-state energy of $\nuc{B}{11}$ is $-11.40 \, \mathrm{MeV} $
from the $\alpha+\alpha+t$ threshold compared with the experimental 
value of $-11.20 \, \mathrm{MeV} $.
%
%
\subsection{Energy convergence of the $3/2^{-}$ states}
\par
\begin{figure}[tb]
  \centering
  \includegraphics[width=5.5cm]{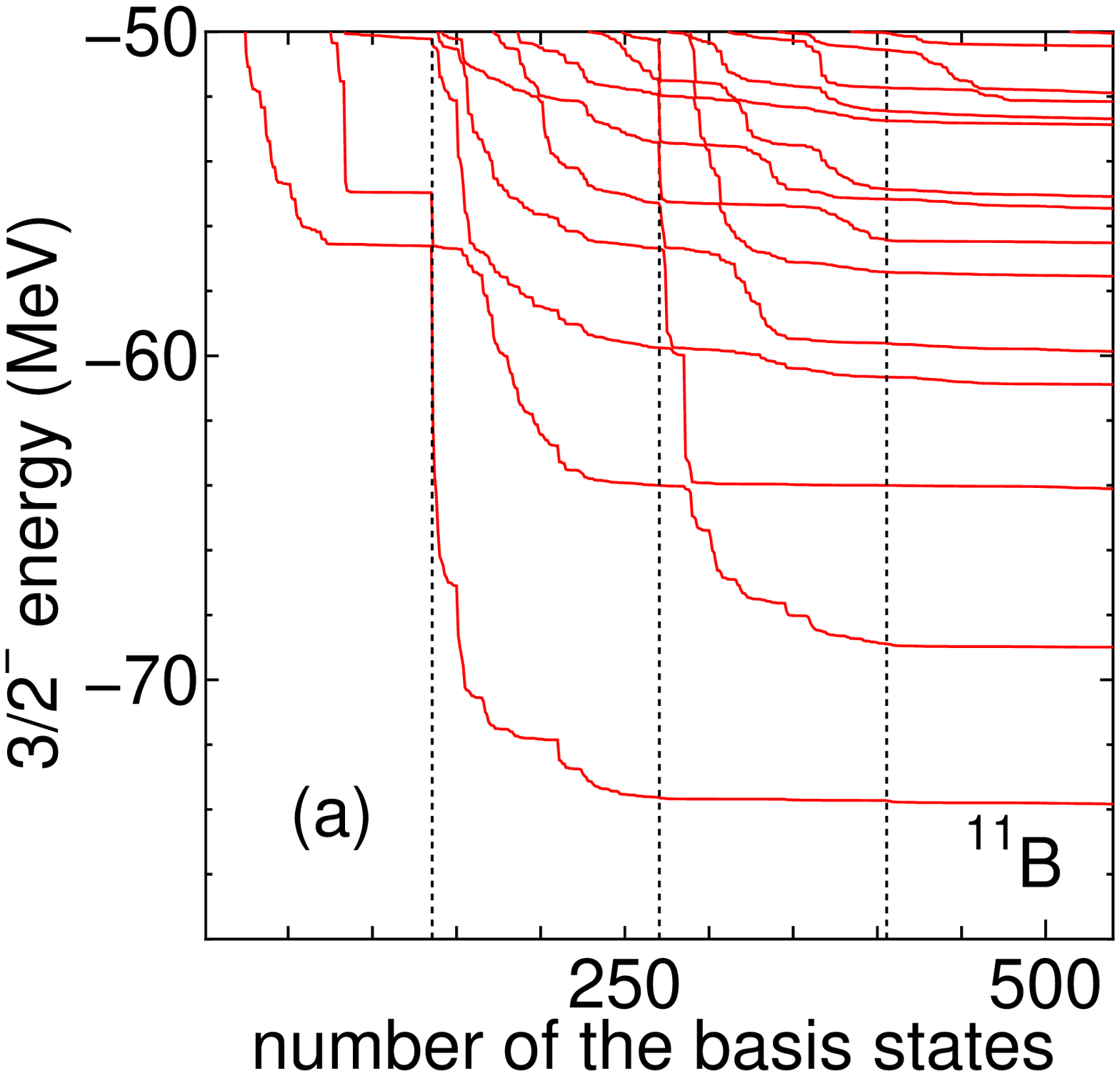} 
  \includegraphics[width=5.5cm]{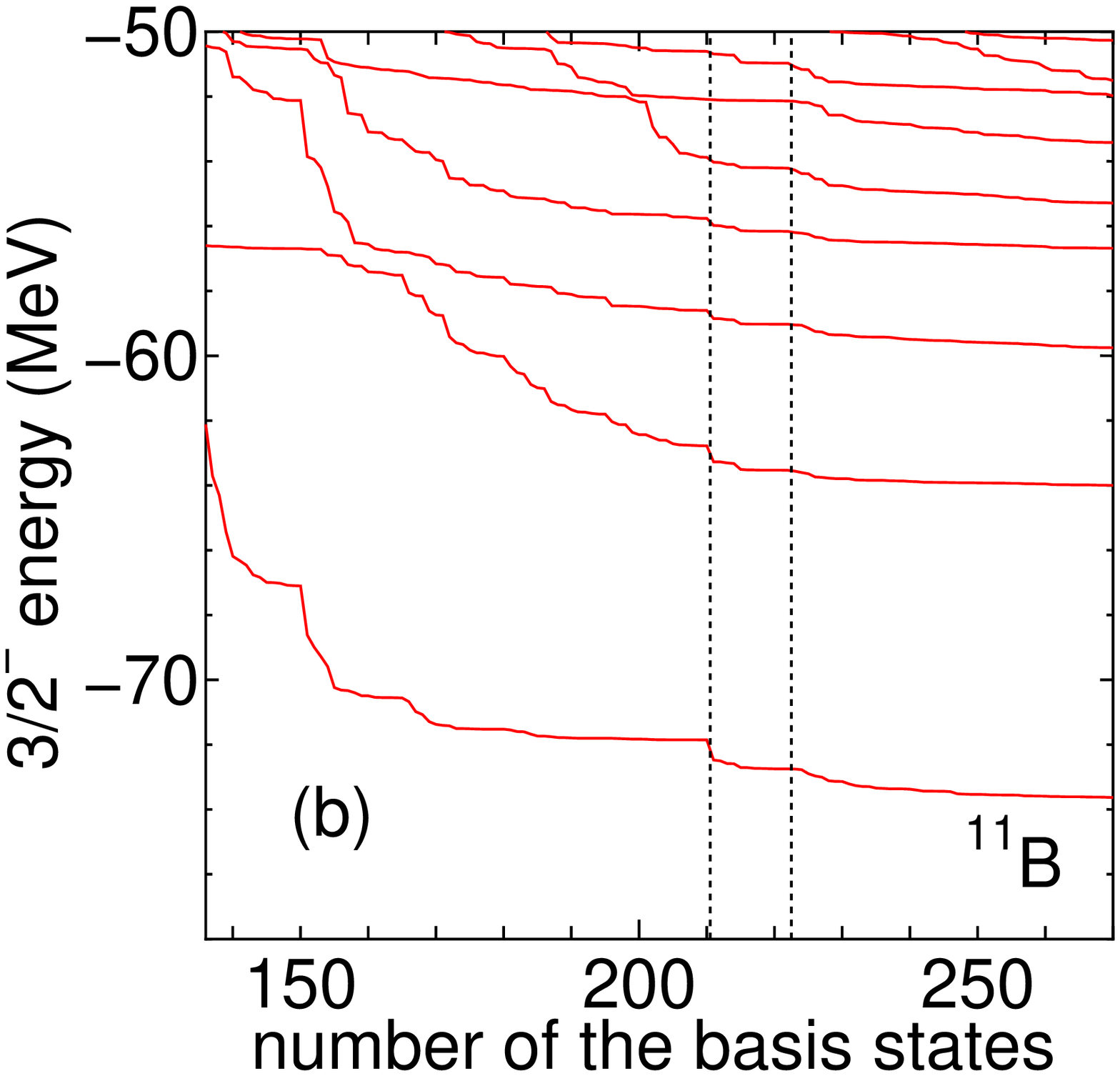} 
  \caption{
    Energy convergence of $ 3/2^{-} $ states of $\nuc{B}{11}{}$.
    The horizontal axis
    shows the number of the basis states.
    (a): 135 basis states are projected to $K=1/2$ (1--135 on the horizontal axis), 
    $K=3/2$ (136--270), $K=-1/2$ (271--405), and $K=-3/2$ (406--540),
    (b): excerpt of the  $K=3/2$ part.}
  \label{eg-conv}
\end{figure}

The energy convergence of the $3/2^{-}$
states of $\nuc{B}{11}$ is shown in Fig.~\ref{eg-conv}(a).
The horizontal axis shows the number of the basis states superposed in the GCM calculation.
As mentioned before, we generated
135 basis states, which are projected to the $J^\pi$ eigenstates with
$K=1/2$, $3/2$, $-1/2$, and $-3/2$,
and thus we diagonalize the norm and Hamiltonian with the dimension of $135 \times 4 = 540$.  
From this figure, the first and third $3/2^{-}$ states are found to have predominantly the 
$K=3/2$ component, whereas the second $3/2^{-}$ state has mainly the $K=-1/2$ component.
Figure~\ref{eg-conv}(b) is the excerpt of the $K=3/2$ part. 
Here the first 75 basis states (136--210 on the horizontal axis) are basis states
with the $\alpha+\alpha+t$ cluster configurations,
the next 12 basis states  (211--222) are AQCM, and the last 48 basis states
(223--270) are
$\alpha+\alpha+3N$ configurations.
\par
From these figures, we can confirm that the inclusion of the breaking effect of 
$\alpha+\alpha+t$ cluster structure contributes to the lowering of the 
ground-state energy by about 2~MeV (for instance, we can compare the energies 
at 210 and 270 on the horizontal axis of  Fig.~\ref{eg-conv}(b)).
The decrease of the energy when cluster breaking is incorporated is about half
compared with $\nuc{C}{12}$, where the spin-orbit effect decreases the ground-state energy
by several MeV (the same interaction is applied to $\nuc{C}{12}$ in Ref.~\cite{PhysRevC.94.064324}).
This is because the spin-orbit acts for a proton in the triton cluster
already within the $\alpha+\alpha+t$ model space;
the antisymmetrization effect excites the triton cluster with the $\left(0s\right)^3$ configuration
to the $p$-shell and the angular momentum projection allows the change of this proton wave function to $p_{3/2}$.
Anyhow, two $\alpha$ clusters and di-neutron cluster in the triton cluster are 
free from the spin-orbit interaction, and breaking these clusters based on AQCM
has a certain effect of decreasing the ground state of about $2\, \mathrm{MeV}$.
%
%
\subsection{Squared overlap with each basis state}
%
\begin{figure}[tb]
  \centering
  \includegraphics[width=5.5cm]{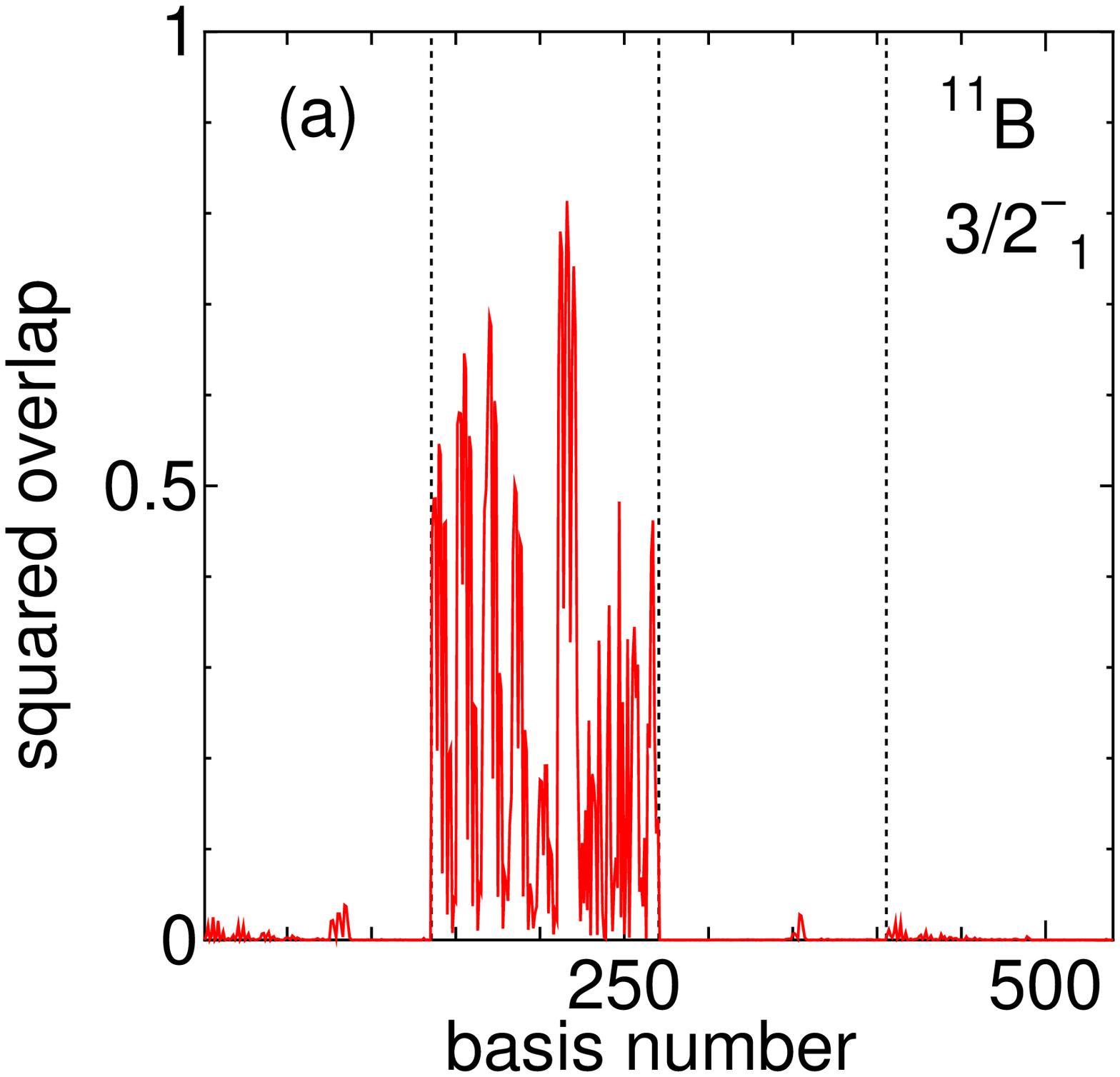} 
  \includegraphics[width=5.5cm]{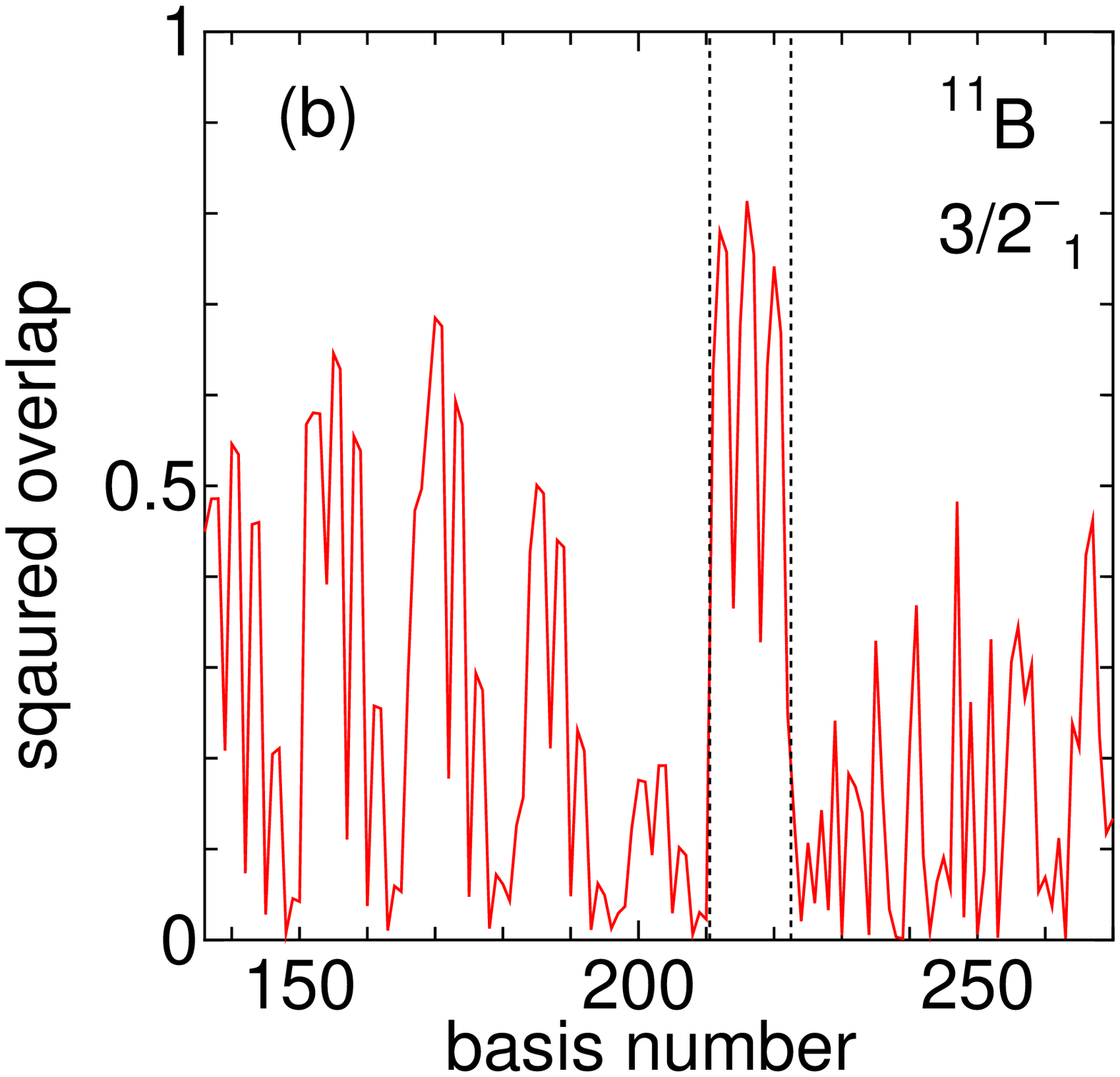} 
  \caption{
    Squared overlap between the ground $3/2^{-}$ state and each basis state.
    (a): basis states with $K=1/2$ (1--135 on the horizontal axis), 
    $K=3/2$ (136--270), $K=-1/2$ (271--405), and $K=-3/2$ (406--540).
    (b): excerpt of the  $K=3/2$ part.}
  \label{sqovlp-1} 
\end{figure}
%
\begin{figure}[tb]
  \centering
  \includegraphics[width=5.5cm]{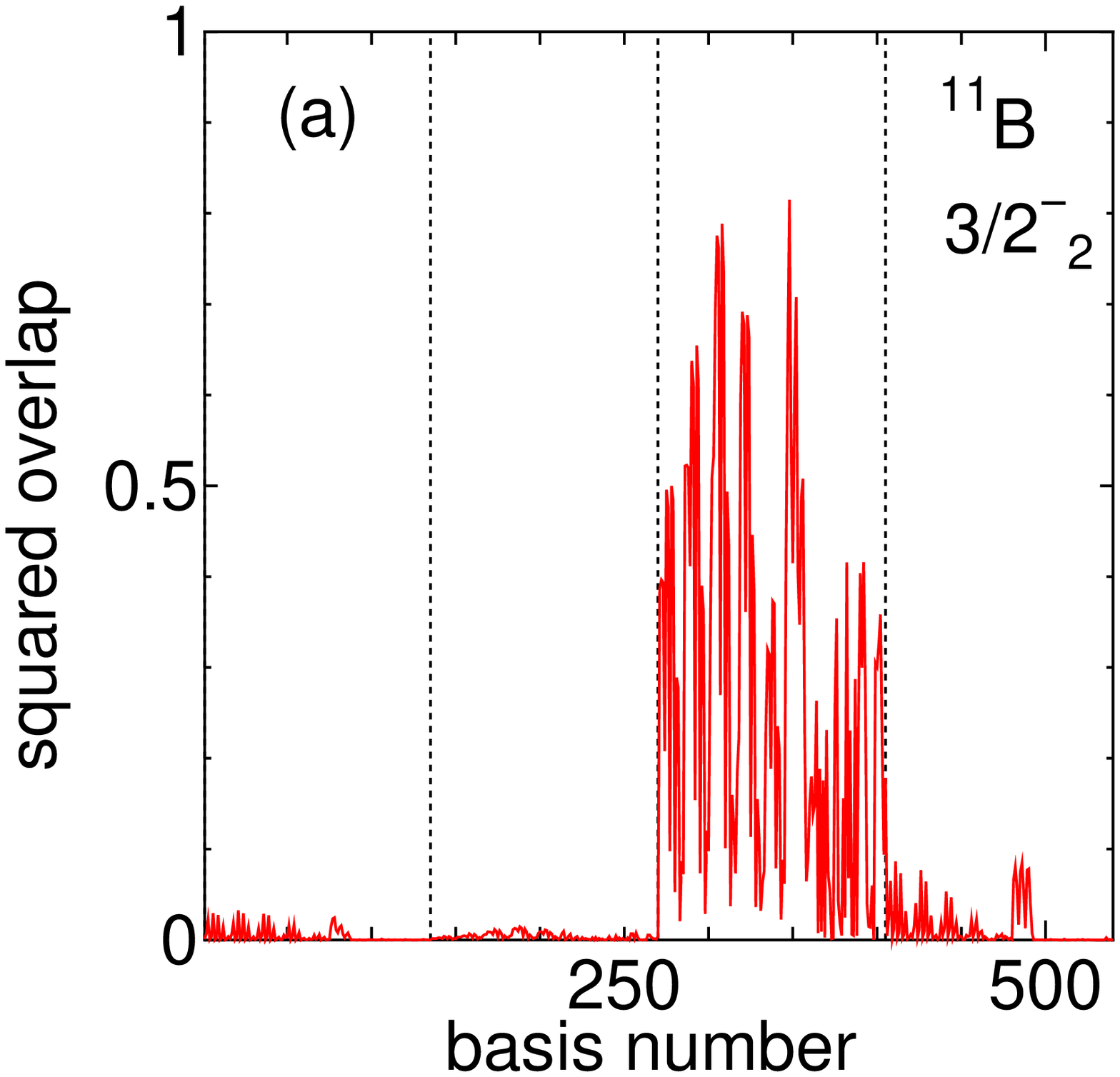} 
  \includegraphics[width=5.5cm]{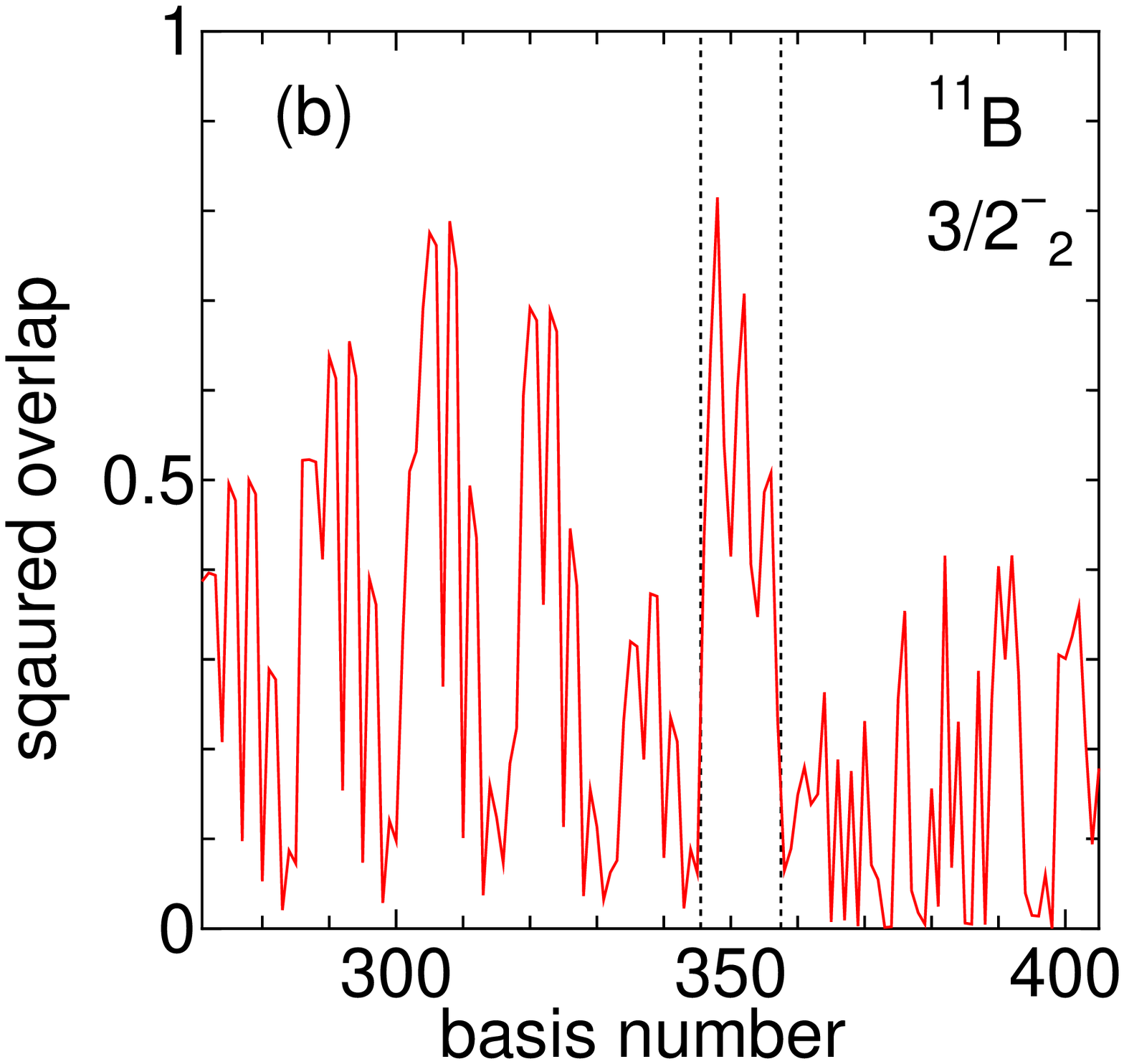} 
  \caption{
    Squared overlap between the second $3/2^{-}$ state and each basis state.
    (a): basis states with $K=1/2$ (1--135 on the horizontal axis), 
    $K=3/2$ (136--270), $K=-1/2$ (271--405), and $K=-3/2$ (406--540).
    (b): excerpt of the  $K=-1/2$ part.}
  \label{sqovlp-2}
\end{figure}
%
\begin{figure}[tb]
  \centering
  \includegraphics[width=5.5cm]{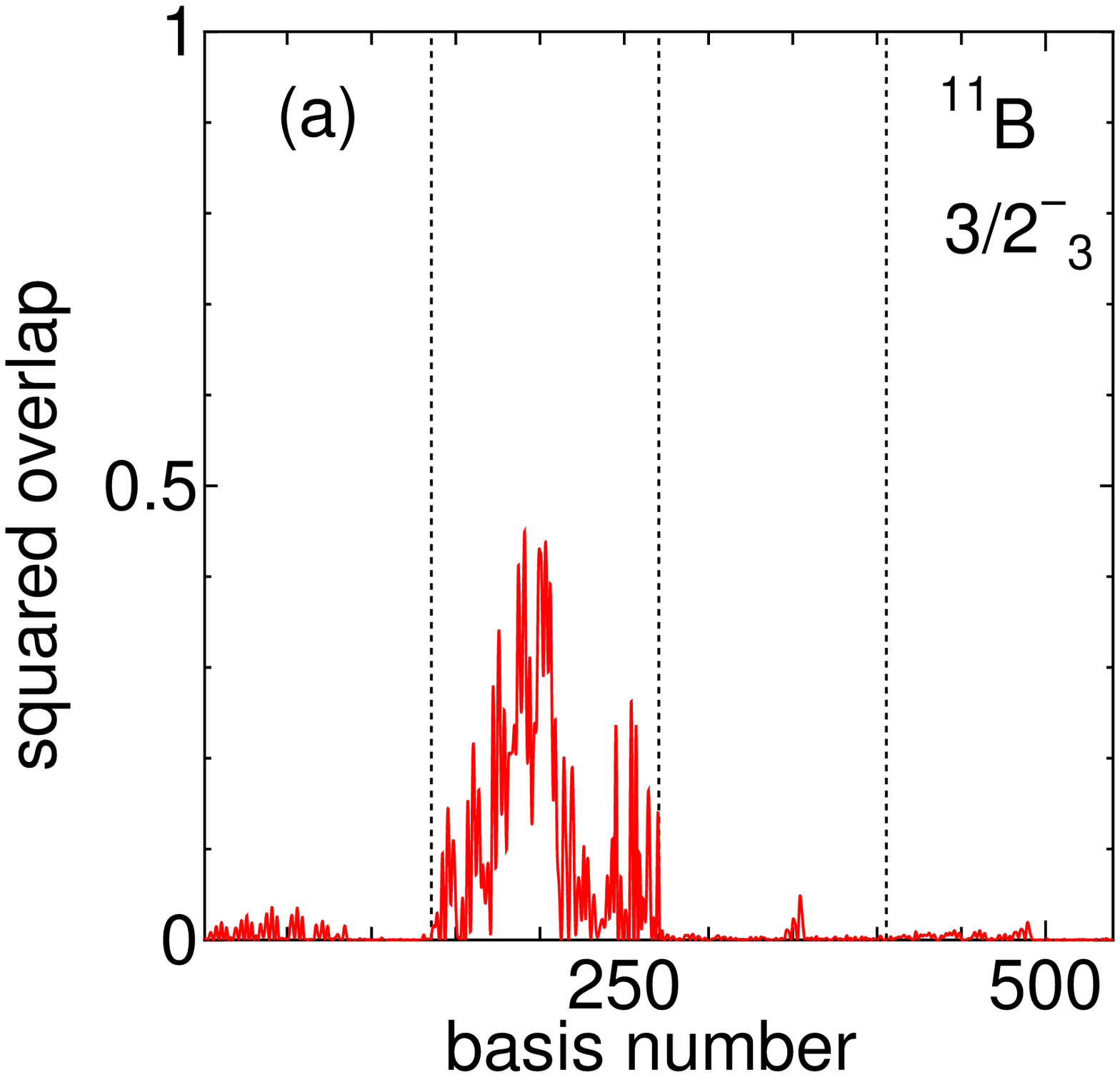} 
  \includegraphics[width=5.5cm]{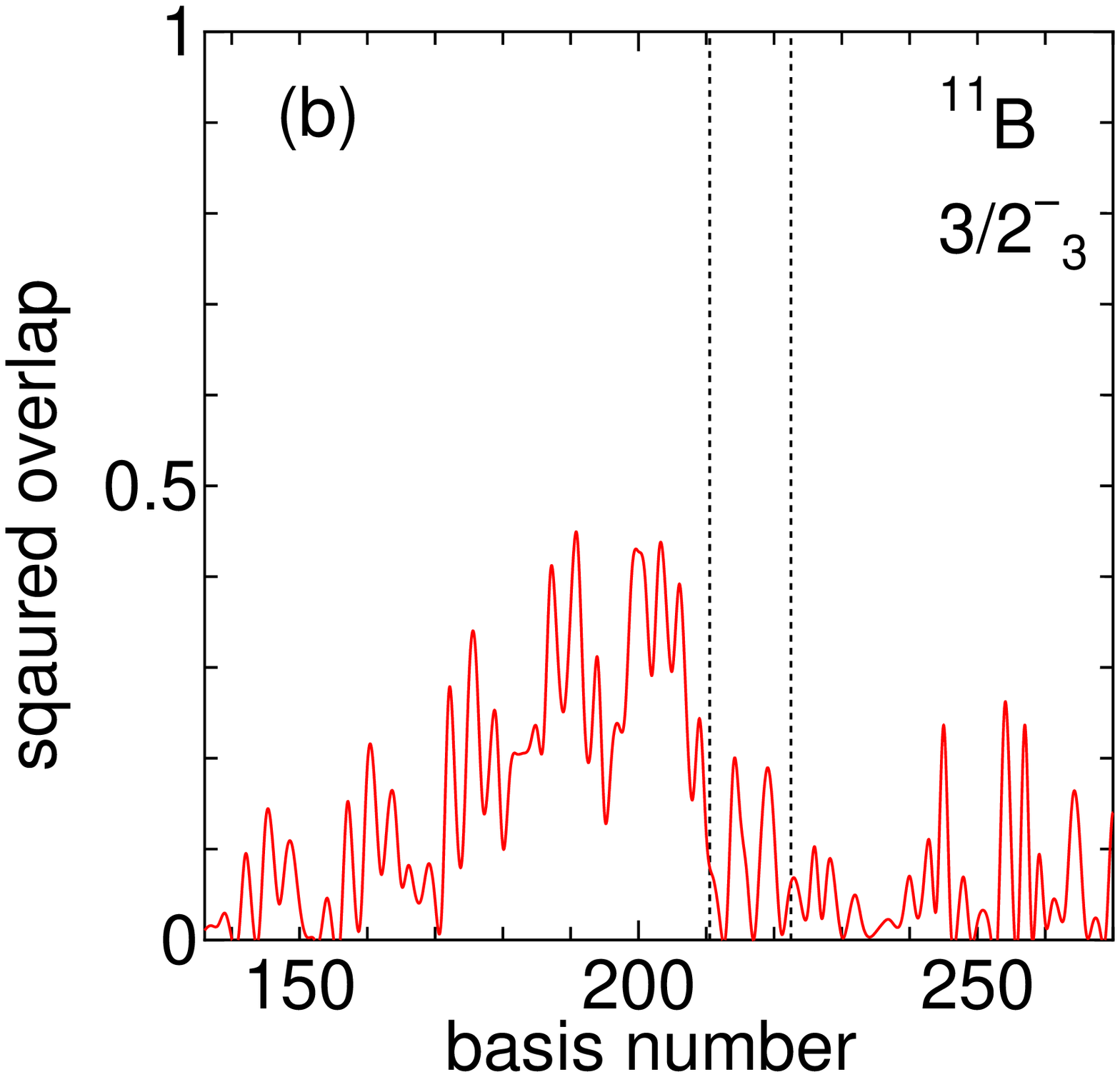} 
  \caption{
    Squared overlap between the third $3/2^{-}$ state and each basis state.
    (a): basis states with $K=1/2$ (1--135 on the horizontal axis), 
    $K=3/2$ (136--270), $K=-1/2$ (271--405), and $K=-3/2$ (406--540).
    (b): excerpt of the  $K=3/2$ part.}
  \label{sqovlp-3}
\end{figure}
\par
Figure~\ref{sqovlp-1}(a)
shows the squared overlap between the ground $3/2^{-}$ state and each basis state.
Here, basis states with $K=1/2$ correspond to 1--135 on the horizontal axis, and
$K=3/2$, $K=-1/2$, and $K=-3/2$ correspond to 136--270, 271--405,
and 406--540 on the horizontal axis. As expected, the contribution of $K=3/2$
is dominant in the ground state.
\par
Figure~\ref{sqovlp-1}(b) is the excerpt of the  $K=3/2$ part. 
As in the previous figure, the first 75 basis states (136--210 on the horizontal axis) are basis states
with the $\alpha+\alpha+t$ cluster configurations,
the next 12 basis states  (211--222) are AQCM, and the last 48 basis states
(223--270) are
$\alpha+\alpha+3N$ configurations.
For the
$\alpha+\alpha+t$ cluster configurations basis states
(136--210), we see five peak structures, and the basis states are classified into five groups with the interval of 15 basis states
(136--150, 151--165, 166--180, 181--195, and 196--210).
These five groups correspond to the basis states with the $\alpha$--$\alpha$ distances of $1$, $2$, $3$, $4$, and $5\,\mathrm{fm}$,
and as we can see, $\alpha$--$\alpha$ distance of 3~fm contributes the most importantly (166--180).
These 15 basis states are further classified into five groups with the interval of three
(166--168, 169--171, 172--174, 175--177, and 178--180)
corresponding to the position of the triton at $1$, $2$, $3$, $4$, and $5\,\mathrm{fm}$ form the center of $\alpha$--$\alpha$.
The largest squared overlap of $0.68$
within the $\alpha+\alpha+t$ cluster configurations 
is given for the basis state 170 on the horizontal axis,
which corresponds to the $\alpha$--$\alpha$ distance is $3\,\mathrm{fm}$, and the triton is located at $2\,\mathrm{fm}$ from the center
of $\alpha$--$\alpha$ with the angle of $30^\circ$.
Therefore, the ground state is found to have the $\alpha+\alpha+t$ component of about $70\,\%$.
\par
Furthermore, we can get insight about the overlap with the higher excited cluster states of $\nuc{C}{12}$ when a proton is added,
which is important medically, as explained in the introduction.
Such cluster states can be characterized as a state with the three $\alpha$ clusters with the relative distances of $\approx 4 \, \mathrm{fm}$,
since the $\alpha$--$\alpha$ system has the minimum energy with such relative distance in the free space.
In our calculation, the basis states 181--195 correspond to the $\alpha$--$\alpha$ distance of $ 4 \, \mathrm{fm}$,
and in the basis states 190--192,
the last $\alpha$ cluster is placed at 4~fm from the center of two $\alpha$ clusters.
We can confirm that the ground state has the squared overlap of about $50\%$ with those basis states
proving it contains enough amount of seeds for the Hoyle state of $\nuc{C}{12}$.
\par
However, the figures show that the AQCM basis states
(211--222) describe the ground state better than the $\alpha+\alpha+t$ cluster configuration
and they are more important.
These AQCM basis states are classified into three groups with the interval of four, 
and basis states 211--214 are for $\Lambda = 0.1$
(215--218 are for $\Lambda = 0.2$ and 219--222 are for $\Lambda = 0.3$).
The largest squared overlap of 0.81 is  given for the basis state 216 on the horizontal line 
corresponding to the basis states with $\Lambda = 0.2$ and $R = 2 \, \mathrm{fm} $, where $R$ represents the distances between quasi clusters with 
equilateral triangular shape.
The AQCM basis states with finite $\Lambda$ values have larger squared overlaps
than the $\alpha+\alpha+t$ cluster configurations, and this means that
the breaking of clusters due to the spin-orbit contribution is important.
In our previous work on $\nuc{C}{12}$ with the same interaction,
similarly, $\Lambda = 0.2$ was found to give the optimal energy, where the $\alpha$--$\alpha$
distance was reduced to $2\,\mathrm{fm}$ owing to the strong attraction of three quasi clusters.
\par
Figure~\ref{sqovlp-2}(a) shows
the squared overlap between the second $3/2^{-}$ state and each basis state.
The orders of the basis states are the same as in the previous figures,
and it is obvious that $K=-1/2$ is important for this state.
In our basis states, the spin direction of a valence proton is set to spin-up, thus
$K=-1/2$ means that the orbital angular momentum of the proton and its spin are antiparallel,
for which the spin-orbit interaction acts repulsively. 
Figure~\ref{sqovlp-2}(b) is the excerpt of the  $K=-1/2$ part. 
The squared overlaps with the $\alpha+\alpha+t$ cluster states
(271--345 on the horizontal axis) are increases compared with 
the previous cases of the ground state,
and again, the $\alpha$--$\alpha$ distance of 3~fm is found to be most important among them.
However, the squared overlaps with the AQCM basis states (346--357 on the horizontal axis)
have even larger overlaps. Contrary to the ground-state case,
the basis state with $\Lambda = 0.1$ has the largest squared overlap.
\par
Figure~\ref{sqovlp-3}(a) shows
the squared overlap between the third $3/2^{-}$ state,
which has been suggested as a candidate for the cluster state both theoretically and experimentally,  
and each basis state.
Here, we find that the contribution of $K=3/2$ is dominant,
and Figure~\ref{sqovlp-3}(b) is the excerpt of the $K=3/2$ part.
The squared overlaps with the $\alpha+\alpha+t$ cluster states
(136--210 on the horizontal axis) are much larger than 
those of the AQCM basis states (211--222).
We can confirmed that the breaking effect of clusters is small and this state is really a cluster state
as expected.
The $\alpha$--$\alpha$ distance of $4\,\mathrm{fm}$ is
found to have the largest overlap, which can be considered as well developed clustering.
\par
Summarizing, one can find that
$K=3/2$ contributes dominantly to the ground $3/2^{-}$,
where AQCM states dominates the most, especially with $ \Lambda = 0.2 $,
while the $\alpha+\alpha+t$ cluster states,
especially with $\alpha$--$\alpha$ distance $3 \, \mathrm{fm}$,
contributes largely as well;
$K=-1/2$ contributes dominantly to the second $3/2^{-}$,
where AQCM states dominates the most, especially with $ \Lambda = 0.1 $,
while the $\alpha+\alpha+t$ cluster states,
especially with $\alpha$--$\alpha$ distance $3 \, \mathrm{fm}$,
contributes largely as well;
$K=3/2$ contributes dominantly to the third $3/2^{-}$,
where the $\alpha+\alpha+t$ cluster states,
especially with $\alpha$--$\alpha$ distance $4 \, \mathrm{fm}$,
dominates.
\subsection{Squared overlap between $p+\nuc{B}{11}$ and $\nuc{C}{12}$}
\par
\begin{figure}[tb]
  \centering
  \includegraphics[width=5.5cm]{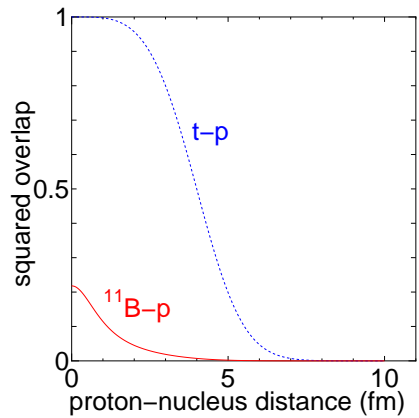} 
  \caption{
    Squared overlap between proton-$\nuc{B}{11}$ and three $\alpha$ clusters
    state as a distance between them (solid line), where the angular momentum is projected to $2^{+}$.
    The dotted line is for the one between proton-triton and $\nuc{He}{4}$ (angular momentum is $0^{+}$).}
  \label{p-nucl-ovlp}
\end{figure}

We have shown how the components of the $\alpha+\alpha+t$ configurations
are mixed in the ground state of $\nuc{B}{11}$. 
Here, finally, we discuss the squared overlap with the three-$\alpha$ cluster configuration
when a proton approaches $\nuc{B}{11}$. 
We start with the case of single-$\alpha$ formation.
The dotted line in Fig.~\ref{p-nucl-ovlp}
shows the squared overlap between an $\alpha$ cluster and $t+p$ as 
a function of the distance between the triton cluster and the proton. 
The total angular momentum is projected to $0^{+}$.
Here the spin orientations of the proton and the one in the triton cluster
are assumed to be anti-parallel. The squared overlap becomes zero
when they are parallel. Therefore, we need an additional factor of $1/2$,
which is not considered in Fig.~\ref{p-nucl-ovlp},
in the real situation of the proton scattering, where spin orientations are not fixed.
The squared overlap between $\alpha$ and $t+p$
is unity at zero-distance and gradually drops as the distance increases.
\par
Next, the solid line shows 
the squared overlap between $p+\nuc{B}{11}$ and three-$\alpha$-cluster
state as a function of the distance between the proton and $\nuc{B}{11}$.
In the real situation of the proton scattering, various three-$\alpha$ cluster states 
of $\nuc{C}{12}$ are created when the proton meets $\nuc{B}{11}$,
and in this calculation, such three-$\alpha$-cluster states are represented by a single configuration
of an equilateral triangular shape with the relative $\alpha$--$\alpha$ distance of $4\,\mathrm{fm}$
(optimal $\alpha$--$\alpha$ distance of $\nuc{Be}{8}$ in the free space).
The target nucleus, $\nuc{B}{11}$, is represented by the most important basis state 
in the previous subsection;
AQCM basis state with $R = 3 \, \mathrm{fm} $ and $\Lambda = 0.2$. To compare with the real situation of the proton scattering,
the target nucleus of $\nuc{B}{11}$ should be angular momentum projected to $3/2^{-}$
and the relative angular momentum between $p$ and $\nuc{B}{11}$ also must be treated as a good quantum number.
However, here these are approximated as the total angular momentum projection,
and the proton is assumed to approach from the perpendicular direction to the three quasi clusters.
The threshold energy of $\nuc{B}{11}+p$ corresponds to $E_x = 15.95668 \, \mathrm{MeV} $ in $\nuc{C}{12}$,
and the resonances slightly above this energy are clinically important sources of the $\alpha$-cluster emission.
According to the database at National Nuclear Data Center, 
the $2^{+}$ resonance state of $\nuc{C}{12}$
at $E_x = 16.10608 \, \mathrm{MeV} $ $\alpha$ decays with the branching ratio of $100\,\%$.
Therefore, here we project the angular momentum to $2^{+}$,
and in this case, the relative wave function
between the proton and $\nuc{B}{11}$ dominantly has the $p$-wave component.
Again the spin orientations of the incident proton and that in the target nucleus are assumed to be
antiparallel, and thus we need an additional factor of $1/2$,
which is not considered in Fig.~\ref{p-nucl-ovlp},
to compare with the real situation.
We can see the squared overlap of $10$--$20\,\%$ at small distances.
The value is not very large, but the proton scattering on $\nuc{B}{11}$ is considered tot have
a sizeable branching ratio for creating the three-$\alpha$ cluster states.
%
%
\section{Conclusions} 
\label{Concl}
\par
We have discussed the
persistence of $\alpha+\alpha+t$ cluster configuration is discussed in the ground state 
of $\nuc{B}{11}$.
The highly excited three-$\alpha$ cluster states of $\nuc{C}{12}$ 
are recently used for cancer treatment.  The proton scattering on 
$\nuc{B}{11}$ creates excited states of $\nuc{C}{12}$ with three $\alpha$ configurations
much above the three-$\alpha$ threshold energy and emitted $\alpha$ particles after the decay destroy cancer cells.
From the nuclear structure point of view,
it would be intriguing to investigate whether $\alpha+\alpha+t$ structure persist 
in the ground state of $\nuc{B}{11}$; this spin-orbit interaction is 
known to work as a driving force to break the
$\alpha$ clusters and
it is considered that the cluster structure is washed out to some extent
as in the case of $\nuc{C}{12}$.
Although this effect is not investigated within the traditional $\alpha$ cluster models,
the antisymmetrized quasi cluster model (AQCM) allows as to
transform cluster model wave functions to $jj$-coupling shell-model ones,
by which the contribution of the spin-orbit interaction can be included. 
All of these basis states are superposed based on
the framework of the generator coordinate method (GCM),
and cluster-shell competition is microscopically investigated.
\par
It is confirmed that the inclusion of the breaking effect of 
$\alpha+\alpha+t$ cluster structure contributes to the lowering of the 
ground-state energy by about $2\,\mathrm{MeV}$.
The decrease of the energy when cluster breaking is incorporated is about half
compared with $\nuc{C}{12}$, where the spin-orbit effect decreases the ground-state energy
by several MeV, and this is because the spin-orbit acts for a proton in the triton cluster
already within the $\alpha+\alpha+t$ model space;
the antisymmetrization effect the triton cluster to the $p$-shell and the angular momentum projection allows the change of this proton wave function to $p_{3/2}$. 
Anyway, two $\alpha$ clusters and di-neutron cluster in the triton cluster are 
free from the spin-orbit interaction, and breaking these clusters based on AQCM
has a certain effect.
Nevertheless, the ground state has the $\alpha+\alpha+t$ component of about $70\,\%$.
Furthermore, we can get insight that $\nuc{B}{11}$ contains the seeds of the Hoyle state of $\nuc{C}{12}$ when a proton is added.
The three-$\alpha$ cluster state in highly excited states
can be characterized as a state with the three $\alpha$ clusters with the relative distances of $\approx 4\,\mathrm{fm} $
(optimal $\alpha$--$\alpha$ distance of $\nuc{Be}{8}$ in the free space).
We can confirm that the ground state of $\nuc{B}{11}$ has a certain squared overlap
($ 10 $--$ 20 \, \% $)
with the state having the $\alpha$--$\alpha$ distance of $ 4 \, \mathrm{fm} $ when a proton approaches.
\par
The third $3/2^{-}$ state
has been suggested as a cluster state both theoretically and experimentally.
The squared overlaps with the $\alpha+\alpha+t$ cluster states are much larger than 
the AQCM basis states.
The $\alpha$--$\alpha$ distance of $ 4 \, \mathrm{fm} $ is
found to have the largest overlap, which can be considered as well developed clustering.
\begin{acknowledgments}
  This work was supported by JSPS KAKENHI Grant Number 19J20543.
  The numerical calculations have been performed using the computer facility of 
  Yukawa Institute for Theoretical Physics,
  Kyoto University (Yukawa-21). 
\end{acknowledgments}
%
\bibliography{b11_bib}

\begin{thebibliography}{32}%
\makeatletter
\providecommand \@ifxundefined [1]{%
 \@ifx{#1\undefined}
}%
\providecommand \@ifnum [1]{%
 \ifnum #1\expandafter \@firstoftwo
 \else \expandafter \@secondoftwo
 \fi
}%
\providecommand \@ifx [1]{%
 \ifx #1\expandafter \@firstoftwo
 \else \expandafter \@secondoftwo
 \fi
}%
\providecommand \natexlab [1]{#1}%
\providecommand \enquote  [1]{``#1''}%
\providecommand \bibnamefont  [1]{#1}%
\providecommand \bibfnamefont [1]{#1}%
\providecommand \citenamefont [1]{#1}%
\providecommand \href@noop [0]{\@secondoftwo}%
\providecommand \href [0]{\begingroup \@sanitize@url \@href}%
\providecommand \@href[1]{\@@startlink{#1}\@@href}%
\providecommand \@@href[1]{\endgroup#1\@@endlink}%
\providecommand \@sanitize@url [0]{\catcode `\\12\catcode `\$12\catcode
  `\&12\catcode `\#12\catcode `\^12\catcode `\_12\catcode `\%12\relax}%
\providecommand \@@startlink[1]{}%
\providecommand \@@endlink[0]{}%
\providecommand \url  [0]{\begingroup\@sanitize@url \@url }%
\providecommand \@url [1]{\endgroup\@href {#1}{\urlprefix }}%
\providecommand \urlprefix  [0]{URL }%
\providecommand \Eprint [0]{\href }%
\providecommand \doibase [0]{https://doi.org/}%
\providecommand \selectlanguage [0]{\@gobble}%
\providecommand \bibinfo  [0]{\@secondoftwo}%
\providecommand \bibfield  [0]{\@secondoftwo}%
\providecommand \translation [1]{[#1]}%
\providecommand \BibitemOpen [0]{}%
\providecommand \bibitemStop [0]{}%
\providecommand \bibitemNoStop [0]{.\EOS\space}%
\providecommand \EOS [0]{\spacefactor3000\relax}%
\providecommand \BibitemShut  [1]{\csname bibitem#1\endcsname}%
\let\auto@bib@innerbib\@empty
\bibitem [{\citenamefont {{Hoyle}}(1954)}]{Hoyle1954}%
  \BibitemOpen
  \bibfield  {author} {\bibinfo {author} {\bibfnamefont {F.}~\bibnamefont
  {{Hoyle}}},\ }\bibfield  {title} {\bibinfo {title} {{On Nuclear Reactions
  Occuring in Very Hot STARS.I. the Synthesis of Elements from Carbon to
  Nickel.}},\ }\href {https://doi.org/10.1086/190005} {\bibfield  {journal}
  {\bibinfo  {journal} {Astrophys. J. Suppl.}\ }\textbf {\bibinfo {volume}
  {1}},\ \bibinfo {pages} {121} (\bibinfo {year} {1954})}\BibitemShut {NoStop}%
\bibitem [{\citenamefont {Freer}\ and\ \citenamefont
  {Fynbo}(2014)}]{FREER20141}%
  \BibitemOpen
  \bibfield  {author} {\bibinfo {author} {\bibfnamefont {M.}~\bibnamefont
  {Freer}}\ and\ \bibinfo {author} {\bibfnamefont {H.~O.~U.}\ \bibnamefont
  {Fynbo}},\ }\bibfield  {title} {\bibinfo {title} {{The Hoyle state in $
  {}^{12} \mathrm{C} $}},\ }\href
  {https://doi.org/https://doi.org/10.1016/j.ppnp.2014.06.001} {\bibfield
  {journal} {\bibinfo  {journal} {Prog. Part. Nucl. Phys.}\ }\textbf {\bibinfo
  {volume} {78}},\ \bibinfo {pages} {1} (\bibinfo {year} {2014})}\BibitemShut
  {NoStop}%
\bibitem [{\citenamefont {Fujiwara}\ \emph {et~al.}(1980)\citenamefont
  {Fujiwara}, \citenamefont {Horiuchi}, \citenamefont {Ikeda}, \citenamefont
  {Kamimura}, \citenamefont {Kat\ifmmode~\bar{o}\else \={o}\fi{}},
  \citenamefont {Suzuki},\ and\ \citenamefont {Uegaki}}]{PTPS.68.29}%
  \BibitemOpen
  \bibfield  {author} {\bibinfo {author} {\bibfnamefont {Y.}~\bibnamefont
  {Fujiwara}}, \bibinfo {author} {\bibfnamefont {H.}~\bibnamefont {Horiuchi}},
  \bibinfo {author} {\bibfnamefont {K.}~\bibnamefont {Ikeda}}, \bibinfo
  {author} {\bibfnamefont {M.}~\bibnamefont {Kamimura}}, \bibinfo {author}
  {\bibfnamefont {K.}~\bibnamefont {Kat\ifmmode~\bar{o}\else \={o}\fi{}}},
  \bibinfo {author} {\bibfnamefont {Y.}~\bibnamefont {Suzuki}},\ and\ \bibinfo
  {author} {\bibfnamefont {E.}~\bibnamefont {Uegaki}},\ }\bibfield  {title}
  {\bibinfo {title} {{Chapter II. Comprehensive Study of Alpha-Nuclei}},\
  }\href@noop {} {\bibfield  {journal} {\bibinfo  {journal} {Prog. Theor. Phys.
  Supplement}\ }\textbf {\bibinfo {volume} {68}},\ \bibinfo {pages} {29}
  (\bibinfo {year} {1980})}\BibitemShut {NoStop}%
\bibitem [{\citenamefont {Tohsaki}\ \emph {et~al.}(2001)\citenamefont
  {Tohsaki}, \citenamefont {Horiuchi}, \citenamefont {Schuck},\ and\
  \citenamefont {R\"opke}}]{PhysRevLett.87.192501}%
  \BibitemOpen
  \bibfield  {author} {\bibinfo {author} {\bibfnamefont {A.}~\bibnamefont
  {Tohsaki}}, \bibinfo {author} {\bibfnamefont {H.}~\bibnamefont {Horiuchi}},
  \bibinfo {author} {\bibfnamefont {P.}~\bibnamefont {Schuck}},\ and\ \bibinfo
  {author} {\bibfnamefont {G.}~\bibnamefont {R\"opke}},\ }\bibfield  {title}
  {\bibinfo {title} {{Alpha Cluster Condensation in $^{12}C$ and $^{16}O$}},\
  }\href {https://doi.org/10.1103/PhysRevLett.87.192501} {\bibfield  {journal}
  {\bibinfo  {journal} {Phys. Rev. Lett.}\ }\textbf {\bibinfo {volume} {87}},\
  \bibinfo {pages} {192501} (\bibinfo {year} {2001})}\BibitemShut {NoStop}%
\bibitem [{\citenamefont {A.~P.~Cirrone}\ \emph {et~al.}(2018)\citenamefont
  {A.~P.~Cirrone}, \citenamefont {Manti}, \citenamefont {Margarone} \emph
  {et~al.}}]{SciRep8-1141}%
  \BibitemOpen
  \bibfield  {author} {\bibinfo {author} {\bibfnamefont {G.}~\bibnamefont
  {A.~P.~Cirrone}}, \bibinfo {author} {\bibfnamefont {L.}~\bibnamefont
  {Manti}}, \bibinfo {author} {\bibfnamefont {D.}~\bibnamefont {Margarone}},
  \emph {et~al.},\ }\bibfield  {title} {\bibinfo {title} {{First experimental
  proof of Proton Boron Capture Therapy (PBCT) to enhance protontherapy
  effectiveness.}},\ }\href@noop {} {\bibfield  {journal} {\bibinfo  {journal}
  {Sci Rep}\ }\textbf {\bibinfo {volume} {8}},\ \bibinfo {pages} {1141}
  (\bibinfo {year} {2018})}\BibitemShut {NoStop}%
\bibitem [{\citenamefont {Stave}\ \emph {et~al.}(2011)\citenamefont {Stave},
  \citenamefont {Ahmed}, \citenamefont {France}, \citenamefont {Henshaw},
  \citenamefont {M\"{u}ller}, \citenamefont {Perdue}, \citenamefont {Prior},
  \citenamefont {Spraker},\ and\ \citenamefont {Weller}}]{PLB696.26.Stave}%
  \BibitemOpen
  \bibfield  {author} {\bibinfo {author} {\bibfnamefont {S.}~\bibnamefont
  {Stave}}, \bibinfo {author} {\bibfnamefont {M.~W.}\ \bibnamefont {Ahmed}},
  \bibinfo {author} {\bibfnamefont {R.~H.}\ \bibnamefont {France}}, \bibinfo
  {author} {\bibfnamefont {S.~S.}\ \bibnamefont {Henshaw}}, \bibinfo {author}
  {\bibfnamefont {B.}~\bibnamefont {M\"{u}ller}}, \bibinfo {author}
  {\bibfnamefont {B.~A.}\ \bibnamefont {Perdue}}, \bibinfo {author}
  {\bibfnamefont {R.~M.}\ \bibnamefont {Prior}}, \bibinfo {author}
  {\bibfnamefont {M.~C.}\ \bibnamefont {Spraker}},\ and\ \bibinfo {author}
  {\bibfnamefont {H.~R.}\ \bibnamefont {Weller}},\ }\bibfield  {title}
  {\bibinfo {title} {{Understanding the $ {}^{11} \mathrm{B} \left( p, \alpha
  \right) \alpha \alpha $ reaction at the $ 0.675 \, \mathrm{MeV} $
  resonance}},\ }\href {https://doi.org/10.1016/j.physletb.2010.12.015}
  {\bibfield  {journal} {\bibinfo  {journal} {Phys. Lett. B}\ }\textbf
  {\bibinfo {volume} {696}},\ \bibinfo {pages} {26} (\bibinfo {year}
  {2011})}\BibitemShut {NoStop}%
\bibitem [{\citenamefont {Kawabata}\ \emph {et~al.}(2007)\citenamefont
  {Kawabata}, \citenamefont {Akimune}, \citenamefont {Fujita}, \citenamefont
  {Fujita}, \citenamefont {Fujiwara}, \citenamefont {Hara}, \citenamefont
  {Hatanaka}, \citenamefont {Itoh}, \citenamefont {Kanada-En'yo}, \citenamefont
  {Kishi}, \citenamefont {Nakanishi}, \citenamefont {Sakaguchi}, \citenamefont
  {Shimbara}, \citenamefont {Tamii}, \citenamefont {Terashima}, \citenamefont
  {Uchida}, \citenamefont {Wakasa}, \citenamefont {Yasuda}, \citenamefont
  {Yoshida},\ and\ \citenamefont {Yosoi}}]{KAWABATA20076}%
  \BibitemOpen
  \bibfield  {author} {\bibinfo {author} {\bibfnamefont {T.}~\bibnamefont
  {Kawabata}}, \bibinfo {author} {\bibfnamefont {H.}~\bibnamefont {Akimune}},
  \bibinfo {author} {\bibfnamefont {H.}~\bibnamefont {Fujita}}, \bibinfo
  {author} {\bibfnamefont {Y.}~\bibnamefont {Fujita}}, \bibinfo {author}
  {\bibfnamefont {M.}~\bibnamefont {Fujiwara}}, \bibinfo {author}
  {\bibfnamefont {K.}~\bibnamefont {Hara}}, \bibinfo {author} {\bibfnamefont
  {K.}~\bibnamefont {Hatanaka}}, \bibinfo {author} {\bibfnamefont
  {M.}~\bibnamefont {Itoh}}, \bibinfo {author} {\bibfnamefont {Y.}~\bibnamefont
  {Kanada-En'yo}}, \bibinfo {author} {\bibfnamefont {S.}~\bibnamefont {Kishi}},
  \bibinfo {author} {\bibfnamefont {K.}~\bibnamefont {Nakanishi}}, \bibinfo
  {author} {\bibfnamefont {H.}~\bibnamefont {Sakaguchi}}, \bibinfo {author}
  {\bibfnamefont {Y.}~\bibnamefont {Shimbara}}, \bibinfo {author}
  {\bibfnamefont {A.}~\bibnamefont {Tamii}}, \bibinfo {author} {\bibfnamefont
  {S.}~\bibnamefont {Terashima}}, \bibinfo {author} {\bibfnamefont
  {M.}~\bibnamefont {Uchida}}, \bibinfo {author} {\bibfnamefont
  {T.}~\bibnamefont {Wakasa}}, \bibinfo {author} {\bibfnamefont
  {Y.}~\bibnamefont {Yasuda}}, \bibinfo {author} {\bibfnamefont
  {H.}~\bibnamefont {Yoshida}},\ and\ \bibinfo {author} {\bibfnamefont
  {M.}~\bibnamefont {Yosoi}},\ }\bibfield  {title} {\bibinfo {title} {{$ 2
  \alpha + t $ cluster structure in $ {}^{11} \mathrm{B} $}},\ }\href
  {https://doi.org/https://doi.org/10.1016/j.physletb.2006.11.079} {\bibfield
  {journal} {\bibinfo  {journal} {Physics Letters B}\ }\textbf {\bibinfo
  {volume} {646}},\ \bibinfo {pages} {6} (\bibinfo {year} {2007})}\BibitemShut
  {NoStop}%
\bibitem [{\citenamefont {Itagaki}\ \emph {et~al.}(2004)\citenamefont
  {Itagaki}, \citenamefont {Aoyama}, \citenamefont {Okabe},\ and\ \citenamefont
  {Ikeda}}]{PhysRevC.70.054307}%
  \BibitemOpen
  \bibfield  {author} {\bibinfo {author} {\bibfnamefont {N.}~\bibnamefont
  {Itagaki}}, \bibinfo {author} {\bibfnamefont {S.}~\bibnamefont {Aoyama}},
  \bibinfo {author} {\bibfnamefont {S.}~\bibnamefont {Okabe}},\ and\ \bibinfo
  {author} {\bibfnamefont {K.}~\bibnamefont {Ikeda}},\ }\bibfield  {title}
  {\bibinfo {title} {{Cluster-shell competition in light nuclei}},\ }\href
  {https://doi.org/10.1103/PhysRevC.70.054307} {\bibfield  {journal} {\bibinfo
  {journal} {Phys. Rev. C}\ }\textbf {\bibinfo {volume} {70}},\ \bibinfo
  {pages} {054307} (\bibinfo {year} {2004})}\BibitemShut {NoStop}%
\bibitem [{\citenamefont {Mayer}\ and\ \citenamefont {Jensen}(1955)}]{Mayer}%
  \BibitemOpen
  \bibfield  {author} {\bibinfo {author} {\bibfnamefont {M.~G.}\ \bibnamefont
  {Mayer}}\ and\ \bibinfo {author} {\bibfnamefont {H.~G.}\ \bibnamefont
  {Jensen}},\ }\href@noop {} {\bibfield  {journal} {\bibinfo  {journal}
  {``Elementary theory of nuclear shell structure", John Wiley, Sons, New York,
  Chapman, Hall, London}\ } (\bibinfo {year} {1955})}\BibitemShut {NoStop}%
\bibitem [{\citenamefont {Itagaki}\ \emph {et~al.}(2005)\citenamefont
  {Itagaki}, \citenamefont {Masui}, \citenamefont {Ito},\ and\ \citenamefont
  {Aoyama}}]{PhysRevC.71.064307}%
  \BibitemOpen
  \bibfield  {author} {\bibinfo {author} {\bibfnamefont {N.}~\bibnamefont
  {Itagaki}}, \bibinfo {author} {\bibfnamefont {H.}~\bibnamefont {Masui}},
  \bibinfo {author} {\bibfnamefont {M.}~\bibnamefont {Ito}},\ and\ \bibinfo
  {author} {\bibfnamefont {S.}~\bibnamefont {Aoyama}},\ }\bibfield  {title}
  {\bibinfo {title} {{Simplified modeling of cluster-shell competition}},\
  }\href {https://doi.org/10.1103/PhysRevC.71.064307} {\bibfield  {journal}
  {\bibinfo  {journal} {Phys. Rev. C}\ }\textbf {\bibinfo {volume} {71}},\
  \bibinfo {pages} {064307} (\bibinfo {year} {2005})}\BibitemShut {NoStop}%
\bibitem [{\citenamefont {Masui}\ and\ \citenamefont
  {Itagaki}(2007)}]{PhysRevC.75.054309}%
  \BibitemOpen
  \bibfield  {author} {\bibinfo {author} {\bibfnamefont {H.}~\bibnamefont
  {Masui}}\ and\ \bibinfo {author} {\bibfnamefont {N.}~\bibnamefont
  {Itagaki}},\ }\bibfield  {title} {\bibinfo {title} {Simplified modeling of
  cluster-shell competition in carbon isotopes},\ }\href
  {https://doi.org/10.1103/PhysRevC.75.054309} {\bibfield  {journal} {\bibinfo
  {journal} {Phys. Rev. C}\ }\textbf {\bibinfo {volume} {75}},\ \bibinfo
  {pages} {054309} (\bibinfo {year} {2007})}\BibitemShut {NoStop}%
\bibitem [{\citenamefont {Yoshida}\ \emph {et~al.}(2009)\citenamefont
  {Yoshida}, \citenamefont {Itagaki},\ and\ \citenamefont
  {Otsuka}}]{PhysRevC.79.034308}%
  \BibitemOpen
  \bibfield  {author} {\bibinfo {author} {\bibfnamefont {T.}~\bibnamefont
  {Yoshida}}, \bibinfo {author} {\bibfnamefont {N.}~\bibnamefont {Itagaki}},\
  and\ \bibinfo {author} {\bibfnamefont {T.}~\bibnamefont {Otsuka}},\
  }\bibfield  {title} {\bibinfo {title} {Appearance of cluster states in
  $^{13}\mathrm{C}$},\ }\href {https://doi.org/10.1103/PhysRevC.79.034308}
  {\bibfield  {journal} {\bibinfo  {journal} {Phys. Rev. C}\ }\textbf {\bibinfo
  {volume} {79}},\ \bibinfo {pages} {034308} (\bibinfo {year}
  {2009})}\BibitemShut {NoStop}%
\bibitem [{\citenamefont {Itagaki}\ \emph {et~al.}(2011)\citenamefont
  {Itagaki}, \citenamefont {Cseh},\ and\ \citenamefont
  {P\l{}oszajczak}}]{PhysRevC.83.014302}%
  \BibitemOpen
  \bibfield  {author} {\bibinfo {author} {\bibfnamefont {N.}~\bibnamefont
  {Itagaki}}, \bibinfo {author} {\bibfnamefont {J.}~\bibnamefont {Cseh}},\ and\
  \bibinfo {author} {\bibfnamefont {M.}~\bibnamefont {P\l{}oszajczak}},\
  }\bibfield  {title} {\bibinfo {title} {{Simplified modeling of cluster-shell
  competition in $^{20}\mathrm{Ne}$ and $^{24}\mathrm{Mg}$}},\ }\href
  {https://doi.org/10.1103/PhysRevC.83.014302} {\bibfield  {journal} {\bibinfo
  {journal} {Phys. Rev. C}\ }\textbf {\bibinfo {volume} {83}},\ \bibinfo
  {pages} {014302} (\bibinfo {year} {2011})}\BibitemShut {NoStop}%
\bibitem [{\citenamefont {Suhara}\ \emph {et~al.}(2013)\citenamefont {Suhara},
  \citenamefont {Itagaki}, \citenamefont {Cseh},\ and\ \citenamefont
  {P\l{}oszajczak}}]{PhysRevC.87.054334}%
  \BibitemOpen
  \bibfield  {author} {\bibinfo {author} {\bibfnamefont {T.}~\bibnamefont
  {Suhara}}, \bibinfo {author} {\bibfnamefont {N.}~\bibnamefont {Itagaki}},
  \bibinfo {author} {\bibfnamefont {J.}~\bibnamefont {Cseh}},\ and\ \bibinfo
  {author} {\bibfnamefont {M.}~\bibnamefont {P\l{}oszajczak}},\ }\bibfield
  {title} {\bibinfo {title} {Novel and simple description for a smooth
  transition from $\ensuremath{\alpha}$-cluster wave functions to $jj$-coupling
  shell model wave functions},\ }\href
  {https://doi.org/10.1103/PhysRevC.87.054334} {\bibfield  {journal} {\bibinfo
  {journal} {Phys. Rev. C}\ }\textbf {\bibinfo {volume} {87}},\ \bibinfo
  {pages} {054334} (\bibinfo {year} {2013})}\BibitemShut {NoStop}%
\bibitem [{\citenamefont {Itagaki}\ \emph {et~al.}(2016)\citenamefont
  {Itagaki}, \citenamefont {Matsuno},\ and\ \citenamefont
  {Suhara}}]{ptep093D01}%
  \BibitemOpen
  \bibfield  {author} {\bibinfo {author} {\bibfnamefont {N.}~\bibnamefont
  {Itagaki}}, \bibinfo {author} {\bibfnamefont {H.}~\bibnamefont {Matsuno}},\
  and\ \bibinfo {author} {\bibfnamefont {T.}~\bibnamefont {Suhara}},\
  }\bibfield  {title} {\bibinfo {title} {General transformation of $\alpha$
  cluster model wave function to jj-coupling shell model in various 4n
  nuclei},\ }\href@noop {} {\bibfield  {journal} {\bibinfo  {journal} {Prog.
  Theor. Exp. Phys.}\ }\textbf {\bibinfo {volume} {2016}},\ \bibinfo {pages}
  {093D01} (\bibinfo {year} {2016})}\BibitemShut {NoStop}%
\bibitem [{\citenamefont {Matsuno}\ \emph {et~al.}(2017)\citenamefont
  {Matsuno}, \citenamefont {Itagaki}, \citenamefont {Ichikawa}, \citenamefont
  {Yoshida},\ and\ \citenamefont {Kanada-En'yo}}]{ptep063D01}%
  \BibitemOpen
  \bibfield  {author} {\bibinfo {author} {\bibfnamefont {H.}~\bibnamefont
  {Matsuno}}, \bibinfo {author} {\bibfnamefont {N.}~\bibnamefont {Itagaki}},
  \bibinfo {author} {\bibfnamefont {T.}~\bibnamefont {Ichikawa}}, \bibinfo
  {author} {\bibfnamefont {Y.}~\bibnamefont {Yoshida}},\ and\ \bibinfo {author}
  {\bibfnamefont {Y.}~\bibnamefont {Kanada-En'yo}},\ }\bibfield  {title}
  {\bibinfo {title} {Effect of $^{12}\mathrm{C}+\alpha$ clustering on the $e0$
  transition in $^{16}\mathrm{O}$},\ }\href@noop {} {\bibfield  {journal}
  {\bibinfo  {journal} {Prog. Theor. Exp. Phys.}\ }\textbf {\bibinfo {volume}
  {2017}},\ \bibinfo {pages} {063D01} (\bibinfo {year} {2017})}\BibitemShut
  {NoStop}%
\bibitem [{\citenamefont {Matsuno}\ and\ \citenamefont
  {Itagaki}(2017)}]{ptepptx161}%
  \BibitemOpen
  \bibfield  {author} {\bibinfo {author} {\bibfnamefont {H.}~\bibnamefont
  {Matsuno}}\ and\ \bibinfo {author} {\bibfnamefont {N.}~\bibnamefont
  {Itagaki}},\ }\bibfield  {title} {\bibinfo {title} {{Effects of cluster-shell
  competition and BCS-like pairing in 12C}},\ }\href@noop {} {\bibfield
  {journal} {\bibinfo  {journal} {Prog. Theor. Exp. Phys.}\ }\textbf {\bibinfo
  {volume} {2017}},\ \bibinfo {pages} {123D05} (\bibinfo {year}
  {2017})}\BibitemShut {NoStop}%
\bibitem [{\citenamefont {Itagaki}(2016)}]{PhysRevC.94.064324}%
  \BibitemOpen
  \bibfield  {author} {\bibinfo {author} {\bibfnamefont {N.}~\bibnamefont
  {Itagaki}},\ }\bibfield  {title} {\bibinfo {title} {Consistent description of
  $^{12}\mathrm{C}$ and $^{16}\mathrm{O}$ using a finite-range three-body
  interaction},\ }\href {https://doi.org/10.1103/PhysRevC.94.064324} {\bibfield
   {journal} {\bibinfo  {journal} {Phys. Rev. C}\ }\textbf {\bibinfo {volume}
  {94}},\ \bibinfo {pages} {064324} (\bibinfo {year} {2016})}\BibitemShut
  {NoStop}%
\bibitem [{\citenamefont {Itagaki}\ and\ \citenamefont
  {Tohsaki}(2018)}]{PhysRevC.97.014307}%
  \BibitemOpen
  \bibfield  {author} {\bibinfo {author} {\bibfnamefont {N.}~\bibnamefont
  {Itagaki}}\ and\ \bibinfo {author} {\bibfnamefont {A.}~\bibnamefont
  {Tohsaki}},\ }\bibfield  {title} {\bibinfo {title} {Nontrivial origin for the
  large nuclear radii of dripline oxygen isotopes},\ }\href
  {https://doi.org/10.1103/PhysRevC.97.014307} {\bibfield  {journal} {\bibinfo
  {journal} {Phys. Rev. C}\ }\textbf {\bibinfo {volume} {97}},\ \bibinfo
  {pages} {014307} (\bibinfo {year} {2018})}\BibitemShut {NoStop}%
\bibitem [{\citenamefont {Itagaki}\ \emph {et~al.}(2018)\citenamefont
  {Itagaki}, \citenamefont {Matsuno},\ and\ \citenamefont
  {Tohsaki}}]{PhysRevC.98.044306}%
  \BibitemOpen
  \bibfield  {author} {\bibinfo {author} {\bibfnamefont {N.}~\bibnamefont
  {Itagaki}}, \bibinfo {author} {\bibfnamefont {H.}~\bibnamefont {Matsuno}},\
  and\ \bibinfo {author} {\bibfnamefont {A.}~\bibnamefont {Tohsaki}},\
  }\bibfield  {title} {\bibinfo {title} {Explicit inclusion of the spin-orbit
  contribution in the tohsaki-horiuchi-schuck-r\"opke wave function},\ }\href
  {https://doi.org/10.1103/PhysRevC.98.044306} {\bibfield  {journal} {\bibinfo
  {journal} {Phys. Rev. C}\ }\textbf {\bibinfo {volume} {98}},\ \bibinfo
  {pages} {044306} (\bibinfo {year} {2018})}\BibitemShut {NoStop}%
\bibitem [{\citenamefont {Itagaki}\ \emph
  {et~al.}(2020{\natexlab{a}})\citenamefont {Itagaki}, \citenamefont
  {Afanasjev},\ and\ \citenamefont {Ray}}]{PhysRevC.101.034304}%
  \BibitemOpen
  \bibfield  {author} {\bibinfo {author} {\bibfnamefont {N.}~\bibnamefont
  {Itagaki}}, \bibinfo {author} {\bibfnamefont {A.~V.}\ \bibnamefont
  {Afanasjev}},\ and\ \bibinfo {author} {\bibfnamefont {D.}~\bibnamefont
  {Ray}},\ }\bibfield  {title} {\bibinfo {title} {{Possibility of
  $^{14}\mathrm{C}$ cluster as a building block of medium-mass nuclei}},\
  }\href {https://doi.org/10.1103/PhysRevC.101.034304} {\bibfield  {journal}
  {\bibinfo  {journal} {Phys. Rev. C}\ }\textbf {\bibinfo {volume} {101}},\
  \bibinfo {pages} {034304} (\bibinfo {year} {2020}{\natexlab{a}})}\BibitemShut
  {NoStop}%
\bibitem [{\citenamefont {Itagaki}\ \emph
  {et~al.}(2020{\natexlab{b}})\citenamefont {Itagaki}, \citenamefont {Fukui},
  \citenamefont {Tanaka},\ and\ \citenamefont {Kikuchi}}]{PhysRevC.102.024332}%
  \BibitemOpen
  \bibfield  {author} {\bibinfo {author} {\bibfnamefont {N.}~\bibnamefont
  {Itagaki}}, \bibinfo {author} {\bibfnamefont {T.}~\bibnamefont {Fukui}},
  \bibinfo {author} {\bibfnamefont {J.}~\bibnamefont {Tanaka}},\ and\ \bibinfo
  {author} {\bibfnamefont {Y.}~\bibnamefont {Kikuchi}},\ }\bibfield  {title}
  {\bibinfo {title} {{$^{8}\mathrm{He}$ and $^{9}\mathrm{Li}$ cluster
  structures in light nuclei}},\ }\href
  {https://doi.org/10.1103/PhysRevC.102.024332} {\bibfield  {journal} {\bibinfo
   {journal} {Phys. Rev. C}\ }\textbf {\bibinfo {volume} {102}},\ \bibinfo
  {pages} {024332} (\bibinfo {year} {2020}{\natexlab{b}})}\BibitemShut
  {NoStop}%
\bibitem [{\citenamefont {Itagaki}\ and\ \citenamefont
  {Naito}(2021)}]{PhysRevC.103.044303}%
  \BibitemOpen
  \bibfield  {author} {\bibinfo {author} {\bibfnamefont {N.}~\bibnamefont
  {Itagaki}}\ and\ \bibinfo {author} {\bibfnamefont {T.}~\bibnamefont
  {Naito}},\ }\bibfield  {title} {\bibinfo {title} {{Consistent description for
  cluster dynamics and single-particle correlation}},\ }\href
  {https://doi.org/10.1103/PhysRevC.103.044303} {\bibfield  {journal} {\bibinfo
   {journal} {Phys. Rev. C}\ }\textbf {\bibinfo {volume} {103}},\ \bibinfo
  {pages} {044303} (\bibinfo {year} {2021})}\BibitemShut {NoStop}%
\bibitem [{\citenamefont {Nishioka}\ \emph {et~al.}(1979)\citenamefont
  {Nishioka}, \citenamefont {Saito},\ and\ \citenamefont
  {Yasuno}}]{Nishioka-PTP.62.424}%
  \BibitemOpen
  \bibfield  {author} {\bibinfo {author} {\bibfnamefont {H.}~\bibnamefont
  {Nishioka}}, \bibinfo {author} {\bibfnamefont {S.}~\bibnamefont {Saito}},\
  and\ \bibinfo {author} {\bibfnamefont {M.}~\bibnamefont {Yasuno}},\
  }\bibfield  {title} {\bibinfo {title} {{Structure Study of $ 2 \alpha + t $
  System by the Orthogonality Condition Model}},\ }\href
  {https://doi.org/10.1143/PTP.62.424} {\bibfield  {journal} {\bibinfo
  {journal} {Progress of Theoretical Physics}\ }\textbf {\bibinfo {volume}
  {62}},\ \bibinfo {pages} {424} (\bibinfo {year} {1979})}\BibitemShut
  {NoStop}%
\bibitem [{\citenamefont {Descouvemont}(1995)}]{DESCOUVEMONT1995532}%
  \BibitemOpen
  \bibfield  {author} {\bibinfo {author} {\bibfnamefont {P.}~\bibnamefont
  {Descouvemont}},\ }\bibfield  {title} {\bibinfo {title} {{The $ {}^{7}
  \mathrm{Be} \left( \alpha, \gamma \right) {}^{11} \mathrm{C} $ and $ {}^{7}
  \mathrm{Li} \left( \alpha, \gamma \right) {}^{11} \mathrm{B} $ reactions in a
  microscopic three-cluster model}},\ }\href
  {https://doi.org/https://doi.org/10.1016/0375-9474(94)00784-K} {\bibfield
  {journal} {\bibinfo  {journal} {Nuclear Physics A}\ }\textbf {\bibinfo
  {volume} {584}},\ \bibinfo {pages} {532} (\bibinfo {year}
  {1995})}\BibitemShut {NoStop}%
\bibitem [{\citenamefont {Yamada}\ and\ \citenamefont
  {Funaki}(2010)}]{PhysRevC.82.064315}%
  \BibitemOpen
  \bibfield  {author} {\bibinfo {author} {\bibfnamefont {T.}~\bibnamefont
  {Yamada}}\ and\ \bibinfo {author} {\bibfnamefont {Y.}~\bibnamefont
  {Funaki}},\ }\bibfield  {title} {\bibinfo {title}
  {{$\ensuremath{\alpha}+\ensuremath{\alpha}+t$ cluster structures and
  $^{12}\mathrm{C}$(${0}_{2}^{+}$)-analog states in $^{11}\mathrm{B}$}},\
  }\href {https://doi.org/10.1103/PhysRevC.82.064315} {\bibfield  {journal}
  {\bibinfo  {journal} {Phys. Rev. C}\ }\textbf {\bibinfo {volume} {82}},\
  \bibinfo {pages} {064315} (\bibinfo {year} {2010})}\BibitemShut {NoStop}%
\bibitem [{\citenamefont {Zhou}\ and\ \citenamefont
  {Kimura}(2018)}]{PhysRevC.98.054323}%
  \BibitemOpen
  \bibfield  {author} {\bibinfo {author} {\bibfnamefont {B.}~\bibnamefont
  {Zhou}}\ and\ \bibinfo {author} {\bibfnamefont {M.}~\bibnamefont {Kimura}},\
  }\bibfield  {title} {\bibinfo {title} {{$2\ensuremath{\alpha}+t$ cluster
  structure in $^{11}\mathrm{B}$}},\ }\href
  {https://doi.org/10.1103/PhysRevC.98.054323} {\bibfield  {journal} {\bibinfo
  {journal} {Phys. Rev. C}\ }\textbf {\bibinfo {volume} {98}},\ \bibinfo
  {pages} {054323} (\bibinfo {year} {2018})}\BibitemShut {NoStop}%
\bibitem [{\citenamefont {Kanada-En'yo}(2007)}]{PhysRevC.75.024302}%
  \BibitemOpen
  \bibfield  {author} {\bibinfo {author} {\bibfnamefont {Y.}~\bibnamefont
  {Kanada-En'yo}},\ }\bibfield  {title} {\bibinfo {title} {{Negative parity
  states of $^{11}\mathrm{B}$ and $^{11}\mathrm{C}$ and the similarity with
  $^{12}\mathrm{C}$}},\ }\href {https://doi.org/10.1103/PhysRevC.75.024302}
  {\bibfield  {journal} {\bibinfo  {journal} {Phys. Rev. C}\ }\textbf {\bibinfo
  {volume} {75}},\ \bibinfo {pages} {024302} (\bibinfo {year}
  {2007})}\BibitemShut {NoStop}%
\bibitem [{\citenamefont {Suhara}\ and\ \citenamefont
  {Kanada-En'yo}(2012)}]{PhysRevC.85.054320}%
  \BibitemOpen
  \bibfield  {author} {\bibinfo {author} {\bibfnamefont {T.}~\bibnamefont
  {Suhara}}\ and\ \bibinfo {author} {\bibfnamefont {Y.}~\bibnamefont
  {Kanada-En'yo}},\ }\bibfield  {title} {\bibinfo {title} {{Cluster structures
  in ${}^{11}$B}},\ }\href {https://doi.org/10.1103/PhysRevC.85.054320}
  {\bibfield  {journal} {\bibinfo  {journal} {Phys. Rev. C}\ }\textbf {\bibinfo
  {volume} {85}},\ \bibinfo {pages} {054320} (\bibinfo {year}
  {2012})}\BibitemShut {NoStop}%
\bibitem [{\citenamefont {Tohsaki}(1994)}]{PhysRevC.49.1814}%
  \BibitemOpen
  \bibfield  {author} {\bibinfo {author} {\bibfnamefont {A.}~\bibnamefont
  {Tohsaki}},\ }\bibfield  {title} {\bibinfo {title} {{New effective
  internucleon forces in microscopic \ensuremath{\alpha}-cluster model}},\
  }\href {https://doi.org/10.1103/PhysRevC.49.1814} {\bibfield  {journal}
  {\bibinfo  {journal} {Phys. Rev. C}\ }\textbf {\bibinfo {volume} {49}},\
  \bibinfo {pages} {1814} (\bibinfo {year} {1994})}\BibitemShut {NoStop}%
\bibitem [{\citenamefont {Tamagaki}(1968)}]{PTP.39.91}%
  \BibitemOpen
  \bibfield  {author} {\bibinfo {author} {\bibfnamefont {R.}~\bibnamefont
  {Tamagaki}},\ }\bibfield  {title} {\bibinfo {title} {{Potential Models of
  Nuclear Forces at Small Distances}},\ }\href@noop {} {\bibfield  {journal}
  {\bibinfo  {journal} {Prog. Theor. Phys.}\ }\textbf {\bibinfo {volume}
  {39}},\ \bibinfo {pages} {91} (\bibinfo {year} {1968})}\BibitemShut {NoStop}%
\bibitem [{\citenamefont {Brink}(1966)}]{Brink}%
  \BibitemOpen
  \bibfield  {author} {\bibinfo {author} {\bibfnamefont {D.~M.}\ \bibnamefont
  {Brink}},\ }\href@noop {} {\bibfield  {journal} {\bibinfo  {journal} {Proc.
  Int. School Phys.``Enrico Fermi"}\ }\textbf {\bibinfo {volume} {XXXVI}},\
  \bibinfo {pages} {247} (\bibinfo {year} {1966})}\BibitemShut {NoStop}%
\end{thebibliography}%
%
\end{document}